\documentclass[pdflatex,sn-mathphys-num]{sn-jnl}

\usepackage{graphicx}%
\usepackage{multirow}%
\usepackage{amsmath,amssymb,amsfonts}%
\usepackage{amsthm}%
\usepackage{mathrsfs}%
\usepackage[title]{appendix}%
\usepackage{xcolor}%
\usepackage{textcomp}%
\usepackage{manyfoot}%
\usepackage{booktabs}%
\usepackage{algorithm}%
\usepackage{algorithmicx}%
\usepackage{algpseudocode}%
\usepackage{listings}%
\usepackage{caption,subcaption}

\theoremstyle{thmstyleone}%
\newtheorem{theorem}{Theorem}[section]
\newtheorem{proposition}{Proposition}[section]

\newtheorem{remark}{Remark}[section]
\newtheorem{lemma}{Lemma}[section]
\numberwithin{equation}{section}

\theoremstyle{thmstyletwo}%

\theoremstyle{thmstylethree}%

\def\exp{\mathrm{exp}}

\def\R{\mathbb{R}}
\def\veck{\mathbf{k}}
\def\vecm{\mathbf{m}}
\def\vecv{\mathbf{v}}
\def\vecx{\mathbf{x}}
\def\vecu{\mathbf{u}}
\def\vecy{\mathbf{y}}
\def\vecE{\mathbf{E}}

\def\magB{\mathbf{B}}
\def\maxR{\mathbf{R}}
\def\maxM{\mathbf{M}}
\def\maxQ{\mathbf{Q}}

\newcommand{\revise}[1]{{#1}}
\newcommand{\revR}[1]{{#1}
}

\begin{document}

\title[Control Magnetized Plasma]{Control of a Uniformly Magnetized Plasma with External Electric Fields}

\author*[1]{\fnm{Peiyi} \sur{Chen}}\email{pchen345@wisc.edu}

\author[2]{\fnm{Rogerio} \sur{Jorge}}\email{rogerio.jorge@wisc.edu}
\equalcont{These authors contributed equally to this work.}

\author[1]{\fnm{Qin} \sur{Li}}\email{qinli@math.wisc.edu}
\equalcont{These authors contributed equally to this work.}

\author[1]{\fnm{Yukun} \sur{Yue}}\email{yyue24@math.wisc.edu}
\equalcont{These authors contributed equally to this work.}

\affil*[1]{\orgdiv{Department of Mathematics}, \orgname{University of Wisconsin-Madison}, \orgaddress{\street{480 Lincoln Drive}, \city{Madison}, \postcode{53706}, \state{WI}, \country{USA}}}

\affil[2]{\orgdiv{Department of Physics}, \orgname{University of Wisconsin-Madison}, \orgaddress{\street{1150 University Avenue}, \city{Madison}, \postcode{53706}, \state{WI}, \country{USA}}}


\abstract{Stabilizing plasma dynamics through externally applied electric and magnetic fields is a fundamental control problem. We study this question for a plasma evolving under a uniform external magnetic field. Although the governing dynamics are nonlinear, a linear analysis based on the Laplace–Fourier transform yields actionable insight. In particular, by controlling the location of the roots of the dispersion relation, we propose a general control strategy that restores stability, with the free-streaming solution recovered as a special case. Numerical experiments for Gaussian equilibria and for the Dory–Guest–Harris instability show that the proposed control suppresses the unstable modes and stabilizes the dynamics, in agreement with our theoretical predictions.}


\maketitle

\section{Introduction}\label{sec:intro}

Plasma is often referred to as the fourth state of matter: an ionized gas consisting of free electrons and ions. It is ubiquitous in nature and plays a central role in astrophysical phenomena, and it also underlies the promise of controlled nuclear fusion. \revise{However, plasma dynamics can be highly unstable. Near certain equilibria, even small perturbations may trigger rapid growth of instabilities and drive the system into strongly nonlinear, possibly turbulent, regimes, posing major challenges for both experiments and applications~\cite{Solovev-Shafranov-1995-mag-confinement, zohm2015magnetohydrodynamic, Freidberg_2007}. In fusion devices, for instance, one aims to maintain the plasma close to a stable equilibrium for as long as possible so that the energy remains primarily in kinetic form. Instabilities can instead induce a fast transfer of energy into electric fields and other channels, making energy extraction difficult. Motivated by these considerations, we study plasma control strategies aimed at stabilizing the dynamics near equilibrium.}

\revR{In most experimental configurations, such as tokamaks~\cite{Wesson_tokamak_2004}, stellarators, and mirror machines, plasma is confined by an externally applied magnetic field.} \revise{We focus on a regime in which the magnetic field lines are approximately straight, a setting exemplified by the LAPD (LArge Plasma Device) at UCLA~\cite{Rogers-Ricci-LAPD-2010}. In this case, the plasma dynamics can be modeled by the Vlasov–Poisson (VP) system under a uniform external magnetic field.} Denoting the electron distribution $F(t,\vecx,\vecv)$, the self-generated electric potential $V(t,\vecx)$ and the associated self-generated electric field $\vecE(t,\vecx)$, VP system writes:
\begin{equation} \label{eqn: magnet VP}
\begin{aligned}
\frac{dF}{dt} = \frac{\partial F}{\partial t} + \vecv\cdot \nabla_{\vecx} F + (\vecE+ \vecv \times \magB)\cdot \nabla_{\vecv} F &= 0\,, \\
\nabla_{\vecx}^2 V &= 1 - \int_{\R^3} F\,d\vecv\,, \\
\vecE &= - \nabla_{\vecx} V\,.
\end{aligned}
\end{equation}
where \revise{$\frac{dF}{dt}$ denotes the total time variation of $F$}, {$\vecx=(x_1,x_2,x_3)^\top \in \mathbb{T}^3, \vecv=(v_1,v_2,v_3)^\top \in\mathbb{R}^3$} are the spatial and velocity coordinates. $\magB = (0, 0, B_0)^\top$ is the fixed external magnetic field pointing in the $x_3$-direction, with $B_0>0$. For notational convenience, throughout the paper, we denote $\vecx_\perp:= (x_1,x_2)^\top, \vecv_\perp:=(v_1,v_2)^\top$. 

\revR{The added external magnetic field brings features to the VP system compared to the one without them~\cite{Schaeffer-1991-vlasov-poisson, PFAFFELMOSER1992281, Mouhot-Villani-2011-landau, Skubachevskii2015-zl, Belyaeva_2020_VP_cylinder, Skubachevskii_2021_VP_confine}}. The foremost discovery is that the structure of equilibrium distributions has changed. Due to the impact of the external magnetic field, only distribution that has radial symmetry in $\vecv_\perp=(v_1,v_2)$ space can serve as equilibrium states. Throughout the paper, we write our equilibrium state as a separable function $\mu(\vecv) = \mu_\perp(\vecv_\perp)\mu_\parallel(v_3)$. As in the classical VP stability analysis, some equilibrium states are stable while others are not, depending on the profile of $\mu_\perp$ and $\mu_\parallel$. In this paper, we place focus on two kinds of equilibrium states with $\mu_\perp$ being either a Gaussian or ring-shaped. The former was shown in~\cite{Bedrossian-Wang-2020-VP} to have non-damping components of electric energy, and the latter can form Dory-Guest-Harris instability~\cite{DGH-1965-PRL, Tataronis-Crawford-1970-perpendicular, VOGMAN2014101}. \revise{Another characteristic feature of the Vlasov–Poisson system in the presence of a uniform magnetic field is the appearance of Bernstein modes~\cite{Bernstein-1958, Sukhorukov-1997-Bernstein, Bedrossian-Wang-2020-VP}. When $\magB$ is directed along the $x_3$-axis, it is natural to distinguish the dynamics parallel and perpendicular to the field. Bernstein modes correspond to electrostatic waves with perpendicular wavevector, i.e. $k_\parallel=0$; in our notation this means $\veck=(k_1,k_2,0)^\top$. This restriction places the focus on stability properties in the transverse plane $\vecx_\perp=(x_1,x_2)$, where magnetization induces behavior distinct from the unmagnetized case.} As a consequence, we confine ourselves to these modes in numerical simulations.

In fusion applications, it is of particular interest to control the stability of plasma, which includes enhancing the solution's stability in the stable regime and suppressing instability in the unstable regime. There are many possible approaches to achieve the task, and one strategy is to impose an external electric field. This amounts to adding an extra electric field $H=-\nabla\Phi$ to~\eqref{eqn: magnet VP}. Upon obtaining the stability analysis for different modes, we then investigate the design principle for developing such external electric fields with the goal of controlling the stability of~\eqref{eqn: magnet VP}.

To understand plasma dynamics, modes with $k_3=0$ are routinely investigated in order to assess the relative importance of drift-waves, the Kelvin-Helmholtz (KH) instability, and sheath physics in driving turbulence and setting its saturation levels~\cite{Rogers-Ricci-LAPD-2010}.
Experimentally~\cite{Horton-etal-PhysPlasma-2005-KH, Perez-etal-PhysPlasma-2006-LAPD},
plasma transport in LAPD due to KH modes with $k_3=0$ is studied using an external biasing of the plasma relative to the chamber wall.
Additionally, in the Mirabelle experiment~\cite{Brochard-etal-PhysPlasma-2005-KH},
strong KH mode activity is obtained for various source profiles.
Theoretically, the approximation $k_3 \ll k_1,k_2$ is typically employed in magnetized plasma systems due to the fact that the strong guide field leads to a separation of scales, with turbulence in $x_1$ and $x_2$ occurring on the gyroradius scale and structures along $x_3$ being on the order of the macroscopic length $L$.
These are the approximations used to model magnetized plasmas using either a fluid framework (e.g., MHD and drift-reduced Braginskii equations) or a kinetic one (e.g., drift-kinetic and gyrokinetic equations).
Finally, besides KH and sheath modes, we mention the  Gradient-driven Drift Coupling (GDC) instability, which, as an essentially $k_3 \simeq 0$ mode, is regarded as a candidate mechanism for turbulence in basic, space, and astrophysical plasmas~\cite{Pueschel-etal-PlasmaPhys-2017-LAPD}.

\revR{A cornerstone to understand plasma dynamics is to derive and analyze the dispersion relation. This is closely tied to the PDE theory on the study of the operators' spectrum. In the context of VP system, a celebrated work on Landau damping was rigorously proved in~\cite{Bedrossian-Masmoudi-Mouhot-2016-Landau}. The dispersion relation for the system with an external magnetic field was studied in~\cite{Bedrossian-Wang-2020-VP}, where the authors investigated the stability of the Bernstein modes for $\mu_\parallel$ being a Gaussian.} Extending from their work, we removed the Gaussian assumption on $\mu$ and achieved a more general dispersion relation.

The plasma control problem has also been extensively studied in the literature. It is particularly prominent in the magnetohydrodynamic (MHD) setting; see, e.g.,~\cite{zohm2015magnetohydrodynamic}. In the kinetic framework, a numerical investigation of the classical VP system in the 1D1V setting using an externally applied electric field was carried out in~\cite{EINKEMMER2024112662}. Related numerical studies include~\cite{Knopf2018-fc, Knopf-Weber-vp-mag-control-2020-bw, Patrik_Knopf_control_PIC, ALBI2025113804}, where control is achieved via externally applied magnetic fields. These numerical works were followed by the theoretical development in~\cite{EINKEMMER-Li-Mouhot-Yue-2025}, which introduced the pole-elimination strategy as a general mechanism for suppressing instability. \revise{Building on this idea, we develop a pole-elimination–based control design for the magnetized Vlasov–Poisson system \eqref{eqn: magnet VP}. Our construction uses a detailed analysis of the associated dispersion relation to derive an explicit feedback term $H$ that removes the unstable spectral contributions (i.e., modes with positive growth rates) and thereby stabilizes the dynamics. While conceptually inspired by~\cite{EINKEMMER-Li-Mouhot-Yue-2025}, the present setting introduces substantial new technical challenges: the analysis is carried out in the full 3D3V phase space and must account for the anisotropic geometry induced by a strong external magnetic field, including the distinguished role of the $x_3$-direction. This leads to new phenomena absent in the unmagnetized case, such as the treatment of ring-type equilibria and the presence of Bernstein modes ($k_3=0$), both of which require additional structural considerations in the control design and stability analysis.} \revR{We also note that this direction differs from work on magnetic confinement~\cite{Skubachevskii2015-zl, Belyaeva_2020_VP_cylinder, Skubachevskii_2021_VP_confine}, which focuses on maintaining spatial confinement, instead of stability, of the plasma in bounded geometries under external fields.}


The paper is organized as follows. In Section~\ref{sec:linear_analysis} we provide linear stability analysis, and derive the Penrose condition and dispersion relation for our system. This provides the baseline for understanding the stability of the system without control. Our control strategies are then provided in Section~\ref{sec:control_external_E}. The Penrose condition reveals the structure of instability, and our control strategies target eliminating the growing modes. One control further recovers the free-streaming solution. Numerical evidences are provided in Section~\ref{sec:numerics}.

\section{Linear analysis}\label{sec:linear_analysis}
In this section, we derive the generalized Penrose condition and the dispersion relation of system~\eqref{eqn: magnet VP} in section~\ref{subsec:penrose_condition_derivation}. This set of analytical derivations provides us with insights into the possibility of plasma generating instability. It will be followed in section~\ref{prop:penrose_cond} by two case studies with two different equilibrium states: Gaussian equilibrium and ring-shaped distribution, in which we apply the previously obtained generalized Penrose condition to evaluate the stability of the system.

Before we conduct linearization, we lay out a few important properties of the system and its equilibrium states in the following proposition.
\begin{proposition} Vlasov-Poisson system with external magnetic field~\eqref{eqn: magnet VP} satisfies the following properties:
\begin{itemize}
\item Any function of the following form is an equilibrium to~\eqref{eqn: magnet VP}:
\begin{align*}
\mu(\vecv) \propto \mu_{\perp}\left(\sqrt{v_1^2 + v_2^2}\right) \mu_{\parallel}(v_3)\,,
\end{align*}
This includes the standard Gaussian distribution or anisotropic Gaussian with different temperatures for $\vecv_\perp$ and $v_3$.
\item The Vlasov-Poisson system~\eqref{eqn: magnet VP} conserves total energy, \[\mathcal{E}_{\text{total}}:=\frac{1}{2}\int |\vecv|^2 F(t,\vecx,\vecv)\,d\vecv d\vecx + \frac{1}{2}\int |\vecE|^2(t,\vecx)\,d\vecx\,,\]
and the second term is the electric energy:
\begin{equation}\label{eqn:electric_energy}
    \mathcal{E}(t):=\frac{1}{2}\int |\vecE|^2(t,\vecx)\,d\vecx.
\end{equation}
\end{itemize}
\end{proposition}

The proof for the proposition is standard material and is omitted from here. It is also worth noting that the $\magB$-component of the acceleration is $\vecv\times \magB$, and this component preserves the magnetic moment $\frac{\vecv_\perp^2}{2B_0}$.

For any given equilibrium $\mu(\vecv)$ of~\eqref{eqn: magnet VP}, we consider a small perturbation around the equilibrium, i.e., $F:=\mu(\vecv)+f$. The perturbation solves the linearized magnetized Vlasov-Poisson system: 
\begin{align}
& \partial_t f + \vecv\cdot \nabla_{\vecx} f + (v_2 B_0 \partial_{v_1} f - v_1 B_0\partial_{v_2} f) + \vecE\cdot \nabla_{\vecv} \mu(\vecv) = 0\,, \quad
\revise{\vecx\in \mathbb{T}^3, \vecv\in\mathbb{R}^3}
\label{eqn: linearized VP}\\
& \nabla_{\vecx}^2 U = -\int_{\R^3} f\,d\vecv\,, \label{eqn: linearized U}\\
& \vecE = - \nabla_{\vecx} U\,. \label{eqn: linearized E}
\end{align}
The system is equipped with initial data $f(t=0,\cdot,\cdot)=f_0$. It is on this linearized equation that we conduct studies on the dispersion relation.

\subsection{Stability condition.\label{subsec:penrose_condition_derivation} }

We first perform the derivation of the dispersion relation for~\eqref{eqn: linearized VP}. The technique has been deployed in~\cite{Bedrossian-Masmoudi-Mouhot-2016-Landau} for Vlasov-Poisson, a slightly simpler system.

To do so, we are to denote the Fourier transform of $f(t,\vecx,\vecv)$ of phase space as
\[
\hat{f}(t,\veck,\vecm):=\int_{\mathbb{T}^3}\int_{\mathbb{R}^3} e^{-i\veck\cdot\vecx - i \vecm\cdot\vecv} f(t,\vecx,\vecv)\,d\vecx d\vecv\,, \veck\in\mathbb{Z}^3\,,\vecm\in \mathbb{R}^3\,,
\]
and the Laplace transform $f(t)$ in time $t$ as
\[
L[f](\lambda):=\int_0^{\infty} e^{-\lambda t} f(t)\,dt\,, \lambda \in \mathbb{C}\,.
\]

The dispersion relation is summarized in the following proposition.
\begin{proposition}[Penrose condition]\label{prop:penrose_cond}
The VP system~\eqref{eqn: linearized VP} in Laplace-Fourier space translates to:
\begin{equation}\label{eqn:solvability}
P(\veck,\lambda)L[\hat{U}] = L[\hat{S}] \,,\quad\forall \veck\in\mathbb{Z}^3\,,\lambda\in\mathbb{C}
\end{equation}
where $P(\veck,\lambda)$ is:
\begin{align}\label{eq:penrose_Pk_lam}
P(\veck,\lambda) = \veck\cdot(I+L[\hat{\mu}\tilde{\maxR}](\lambda, \veck)) \veck\,,
\end{align}
$U$ is the self-generating electric potential, and
\begin{itemize}
    \item $S$ is the free streaming density defined as:
\begin{equation}\label{eq:S_source_free_sol}
S(t,\vecx) = \int f_{0}(\vecy^{-1}(t;\vecx,\vecv),\vecu^{-1}(t;\vecv)) d\vecv\,,
\end{equation}
where $(\vecy(t;\vecx,\vecv),\vecu(t;\vecv))$ denotes the position and velocity of a particle following characteristics of the VP system with initial data at $(\vecx,\vecv)$, see explicit calculation in~\eqref{eqn:y_u_explicit}, and $f_0$ denotes the initial condition.
\item 
\begin{equation}\label{eqn:R_tilde}
\begin{aligned}
\tilde{\maxR}(t) := \begin{pmatrix} 
\dfrac{1}{B_0} \left(\maxR\left(\frac{\pi}{2}-B_0t\right) - \maxR\left(\frac{\pi}{2}\right)\right) & 0 \\
0 & t
\end{pmatrix}\,,\quad\text{where}\quad \maxR(\alpha) :=\begin{pmatrix} \cos(\alpha) & \sin(\alpha) \\ 
-\sin(\alpha) & \cos(\alpha) \end{pmatrix}\,.
\end{aligned}
\end{equation}
\end{itemize}

Furthermore, if there exists some $\kappa_0>0$ that
\begin{align}\label{eqn: penrose cond}
\inf_{\veck\in\mathbb{Z}^3, \mathrm{Re}(\lambda)>0}\, \left|\veck\cdot(I+L[\hat{\mu}\tilde{\maxR}](\lambda, \veck)) \veck\right|\geq \kappa_0\,,
\end{align}
then the equilibrium $\mu$ is stable, in the sense that electric energy $\mathcal{E}(t)$ does not blow up. 
\end{proposition}

\begin{remark}
    It is straightforward to see~\eqref{eqn:solvability} is an extension from the VP system without the external field. Indeed, without the $\magB$, one has the same equation that writes:
    \begin{equation}\label{eq:VP_no_B_dispersion}
    P(\veck,\lambda)L[\hat{U}]= L[\hat{S}]\quad\text{with}\quad P(\veck,\lambda) = \veck\cdot(I + L[t\hat{\mu}(\lambda, \veck t)])\veck 
    \end{equation}
    In comparison to~\eqref{eq:VP_no_B_dispersion}, our equation~\eqref{eqn:solvability} has the same right-hand side that is also generated by the free-streaming solution, and our dispersion function $P$ now involves the rotational operator~\eqref{eqn:R_tilde}. This rotation is a natural consequence of the gyro-effect induced by $\magB$.
\end{remark}

\revise{
Complementary to the Penrose condition~\eqref{eqn: penrose cond}, suppose that there exist
$\veck \in \mathbb{Z}^3$ and $\lambda_0 \in \mathbb{C}$ such that $P(\veck,\lambda_0)=0$, then the temporal behavior of the corresponding mode is characterized by the complex root $\lambda_0$.
In particular, the real part of $\lambda_0$, $\mathrm{Re}(\lambda_0)$, determines whether the mode grows, decays, or remains neutral in time.
If $\mathrm{Re}(\lambda_0)=0$, the associated electric energy does not grow or decay and corresponds to a neutrally stable mode.
If $\mathrm{Re}(\lambda_0)>0$, the mode exhibits exponential growth in time, indicating a potential instability.
}

Before proving the proposition, we first prepare the free-streaming solution and characteristics of~\eqref{eqn: linearized VP}. Setting the free-streaming component of the equation 
\begin{equation}\label{eq:free_stream_eq}
\partial_t f_{\mathrm{f}}+\vecv\cdot \nabla_{\vecx} f_{\mathrm{f}} + (\vecv\times \magB) \cdot \nabla_{\vecv} f_{\mathrm{f}}=0\,,
\end{equation}
we notice that 
\begin{align}
\dfrac{d}{dt} \vecy(t) & = \vecu(t)\,,\\
\dfrac{d}{dt} \vecu(t) & = \begin{pmatrix} 
0 & B_0 & 0 \\ -B_0 & 0 & 0 \\ 0 & 0 & 0
\end{pmatrix} \vecu(t)\,.
\end{align}
For this linear ODE system, the solution is explicit \revR{--- the celebrated the Larmour trajectory (rotation)}:
\begin{align}\label{eqn:y_u_explicit}
\begin{pmatrix} \vecu_{\perp}(t)\\ u_3(t)\end{pmatrix} & = 
\begin{pmatrix} \maxR(B_0 t) & 0 \\ 0 & 1 \end{pmatrix}
\begin{pmatrix} \vecv_{\perp}\\ v_3\end{pmatrix}\,,\\
\begin{pmatrix} \vecy_{\perp}(t)\\ y_3(t)\end{pmatrix} & = \begin{pmatrix} \vecx_{\perp}\\ x_3\end{pmatrix} + 
\begin{pmatrix} \maxM(t) + \maxQ\, & 0 \\ 0 & t \end{pmatrix}
\begin{pmatrix} \vecv_{\perp}\\ v_3\end{pmatrix}\,.
\end{align}
where $\maxM$ and $\maxQ$ are auxiliary matrices defined by:
\[
\maxM(t) : = \frac{1}{B_0} \maxR\left(B_0t - \frac{\pi}{2}\right) = \frac{1}{B_0}\begin{pmatrix} \sin(B_0 t) & -\cos(B_0 t) \\ 
\cos(B_0 t) & \sin(B_0 t) \end{pmatrix} \,, \quad
\maxQ: = \frac{1}{B_0}\begin{pmatrix} 0 & 1 \\ -1 & 0 \end{pmatrix}\,.
\]
As a consequence,
\begin{equation}\label{eq:free_stream_sol}
f_{\mathrm{f}}(t,\vecx, \vecv) = f_0(\vecx_\perp - (\maxM(t) + \maxQ) \maxR^{-1}(B_0 t) \vecv_\perp, x_3 - v_3 t, \maxR^{-1}(B_0 t) \vecv_\perp, v_3)\,.
\end{equation}
and
\[
S(t,x)=\int f_{\mathrm{f}}(t;\vecx,\vecv)\, d\vecv\,,
\]
recovering~\eqref{eq:S_source_free_sol}. Moreover, define
\[
g(t,\vecx,\vecv):=f(t,\vecy(t;\vecx, \vecv), \vecu(t;\vecv))\,,
\]
we have, by plugging in \revise{the linearized  equation of $f$~\eqref{eqn: linearized VP}}:
\begin{align}
\dfrac{\mathrm{d}g}{\mathrm{d}t}(t,\vecx,\vecv) & = \partial_t f + \dfrac{d\vecy}{dt}\cdot \nabla_{\vecx} f + \dfrac{d\vecu}{dt}\cdot \nabla_{\vecv} f \,, \nonumber \\
& = \partial_t f + \vecu(t)\cdot \nabla_{\vecx} f + B_0 u_2 \partial_{v_1} f - B_0 u_1 \partial_{v_2} f\, \nonumber \\
& = - \vecE(\vecy(t;\vecx,\vecv))\cdot \nabla_{\vecv} \mu(\vecu(t;\vecv))\,. \label{eqn: FT-g-E-mu}
\end{align}

Proposition~\ref{prop:penrose_cond} is proved by conducting a Laplace-Fourier transform of equation~\eqref{eqn: FT-g-E-mu}.

\begin{proof}[Proof of Proposition~\ref{prop:penrose_cond}] 
We first notice there are two equations connecting $g$ and $\vecE$, one by~\eqref{eqn: FT-g-E-mu}, and the second by the definition of electric field:
\begin{align*}
\nabla_{\vecx} \cdot \vecE(t,\vecx) & = \int_{\R^3} f(t,\vecx,\vecv)\,d\vecv = \rho(t,\vecx)\,.
\end{align*}
We need to flip both equations to the Fourier side. The Fourier transform of the Poisson equation is straightforward. By definition, 
\begin{align}
f(t,\vecx,\vecv) = g(t; \vecx_\perp - (\maxM(t) + \maxQ) \maxR(B_0 t)^{-1} \vecv_\perp, x_3 - v_3t, \maxR(B_0 t)^{-1} \vecv_\perp, v_3)\,.
\end{align}
So
\begin{align}
i \veck\cdot \hat{E}(t,\veck) & = \hat{\rho}(t,\veck) = \int_{\mathbb{T}^3} \int_{\R^3} e^{-i\veck\cdot \vecx} f(t,\vecx,\vecv)\,d\vecv d\vecx\,,\nonumber \\
& = \int_{\mathbb{T}^3}\int_{\R^3} e^{-i\veck\cdot \vecx} g(t; \vecx_\perp - (\maxM(t) + \maxQ) \maxR(B_0 t)^{-1} \vecv_\perp, x_3 - v_3t, \maxR(B_0 t)^{-1} \vecv_\perp, v_3)\,d\vecv d\vecx\,, \nonumber \\
&  = \int_{\mathbb{T}^3}\int_{\R^3} e^{-i\veck_{\perp}\cdot (\vecx_{\perp}- (\maxM(t) + \maxQ) \maxR(B_0 t)^{-1} \vecv_\perp)} e^{-i k_3 (x_3-v_3 t)} e^{-i (\maxM(t) + \maxQ) \maxR(B_0 t)^{-1} \vecv_\perp} e^{-i k_3 t v_3}\nonumber \\
& \hspace{0.75in} g(t; \vecx_\perp - (\maxM(t) + \maxQ) \maxR(B_0 t)^{-1} \vecv_\perp, x_3 - v_3t, \maxR(B_0 t)^{-1} \vecv_\perp, v_3) \,d\vecv d\vecx\,,  \nonumber \\
& = \hat{g}(t, \veck_\perp, k_3, (\maxM(t) + \maxQ)^\top \veck_\perp, k_3 t)\,, \label{eqn: E-FT}
\end{align}
where we used a change of variable.

Similarly, we conduct the Fourier transform on $(\vecx,\vecv)$ of~\eqref{eqn: FT-g-E-mu}. Denoting $\veck$ and $\vecm$ the associated Fourier coefficients, the equation becomes:
\begin{equation} \label{eqn:ghat_time}
\begin{aligned}
\dfrac{\mathrm{d}\hat{g}}{\mathrm{d}t}(t,\veck,\vecm) = & - \int_{\mathbb{T}^3} \int_{\R^3} e^{-i\veck\cdot \vecx - i \vecm\cdot \vecv} \vecE(\vecy(t;\vecx,\vecv)) \cdot \nabla_{\vecv} \mu(\vecu(t;\vecv))\,d\vecx d\vecv\,, \\
= & - \int_{\mathbb{T}^3} \int_{\R^3} e^{-i\veck\cdot \vecx - i\vecm\cdot \vecv} \vecE(\vecx_\perp + (\maxM(t)+\maxQ) \vecv_\perp, x_3 + v_3 t) \cdot \nabla_{\vecv} \mu(\maxR(B_0 t) \vecv_\perp, v_3)\,d\vecx d\vecv\,, \\
= & - \hat{\vecE}(t, \veck) \cdot \int_{\R^3} e^{+i\veck_\perp\cdot (\maxM(t)+\maxQ) \vecv_\perp + ik_3 v_3t - i\vecm\cdot \vecv} \nabla_{\vecv} \mu(\maxR(B_0 t) \vecv_\perp, v_3)\,d\vecx d\vecv\,, \\
= & - i 
\begin{pmatrix}
\hat{\vecE}_\perp(t, \veck) \\ \hat{E}_3(t, \veck) 
\end{pmatrix}
\cdot 
\begin{pmatrix}
\maxR(B_0t) (\vecm_\perp - (\maxM(t) + \maxQ)^\top \veck_\perp)\\
m_3-k_3 t
\end{pmatrix}  \\
& \hspace{1.1in}
\hat{\mu}(\maxR(B_0t)(\vecm_\perp - (\maxM(t) + \maxQ)^\top \veck_\perp), m_3-k_3t)\,.
\end{aligned} 
\end{equation}
This equation holds for all values of $(\veck,\vecm)$. We \revise{choose} a special pair. Let $\vecm=(\vecm_\perp,m_3)$ defined by:
\begin{equation}\label{eqn:FT-m}
\vecm_\perp (\veck):=(\maxM(t) + \maxQ)^\top \veck_\perp\,, \quad
m_3 (\veck) := k_3t\,,
\end{equation} 
then for $0<s<t$,
\begin{align*}
\maxR(B_0 s)(\vecm_\perp (\veck) - (\maxM(s) + \maxQ)^\top \veck_\perp) & = \maxR(B_0 s)(\maxM(t) - \maxM(s))^\top \veck_\perp\\
& = \frac{1}{B_0} \left(\maxR\left(\frac{\pi}{2}-B_0(t-s)\right) - \maxR\left(\frac{\pi}{2}\right)\right) \veck_\perp\,.
\end{align*}
Integrate~\eqref{eqn:ghat_time} in time for this specifically chosen $(\veck,\vecm(\veck))$, and recall the definition of $\tilde{\maxR}$ in~\eqref{eqn:R_tilde}, we have:
\begin{align} \label{eqn: hat g t-s}
& \hat{g}(t, \veck, \vecm(\veck)) - \hat{g}(0, \veck, \vecm(\veck)) +i \int_0^t \begin{pmatrix}
\hat{\vecE}_\perp(s,\veck) \\ \hat{E}_3(s,\veck) \end{pmatrix} 
\cdot \tilde{\maxR}(t-s) 
\begin{pmatrix} \veck_\perp\\ k_3\end{pmatrix} \hat{\mu}(\tilde{\maxR}(t-s)\veck)\,ds = 0\,.
\end{align}

We are going to apply the Laplace transform on all three terms by multiplying $e^{-\lambda t}$ and integrating in $t$-dimension.
\begin{itemize}
    \item[--] By~\eqref{eqn: E-FT}, the first term
    \[
    \hat{g}(t, \veck, \vecm(\veck))=i\veck\cdot \hat{\vecE}(t,\veck)\,.
    \]
    Hence, using the linearity of the Laplace transform operator, the first term becomes: $i\veck\cdot L[\hat{\vecE}]$;
    \item[--] By definition of free-streaming~\eqref{eq:free_stream_sol},
    \(\hat{f}_{\mathrm{f}} = \hat{f}_0(\veck,\vecm(\veck)) = \hat{g}(0,\veck,\vecm(\veck)),\) and with the same calculation as done in~\eqref{eqn: E-FT}, the second term becomes: $L[\hat{S}]$;
    \item[--] To compute the last term, we notice it has the convolution structure between $\hat{\mu}\tilde{\maxR}$ and $\hat{\vecE}$, the Laplace transform thus becomes:
    \begin{align}
& \int_0^\infty \int_0^t e^{-\lambda t} \begin{pmatrix}
\hat{\vecE}_\perp(s, \veck) \\ \hat{E}_3(s, \veck) \end{pmatrix} \cdot \tilde{\maxR}(t-s) 
\begin{pmatrix} \veck_\perp\\ k_3\end{pmatrix} \hat{\mu}(\tilde{\maxR}(t-s) \veck)\,ds dt\,, \nonumber \\
= & \int_0^\infty \int_0^\infty e^{-\lambda w} e^{-\lambda s} \begin{pmatrix}
\hat{\vecE}_\perp(s, \veck) \\ \hat{E}_3(s, \veck) \end{pmatrix} \cdot \tilde{\maxR}(w) 
\begin{pmatrix} \veck_\perp\\ k_3\end{pmatrix} \hat{\mu}(\tilde{\maxR}(w) \veck)\,dw ds\,, \quad (w:=t-s) \,, \nonumber \\
= & \int_0^\infty e^{-\lambda s} \begin{pmatrix}
\hat{E}_\perp(s, \veck) \\ \hat{E}_3(s, \veck) \end{pmatrix} \,ds \cdot 
\int_0^\infty e^{-\lambda t} \tilde{\maxR}(t) \hat{\mu}(\tilde{\maxR}(t) \veck) \,dt \begin{pmatrix} \veck_\perp\\ k_3\end{pmatrix}\\
=& L[\hat{\vecE}] \cdot L[\hat{\mu}\tilde{\maxR}] \veck \,.\label{eq:decouple_LE}
\end{align}
\end{itemize}
Putting these three terms together in~\eqref{eqn: hat g t-s}, we have
\begin{align}\label{eqn: LE system}
i \veck \cdot L[\hat{\vecE}] - L[\hat{S}] + i L[\hat{\vecE}] \cdot L[\hat{\mu}\tilde{\maxR}] \veck & = 0\,.
\end{align}
Noting the relation between $\vecE$ and $U$:
\begin{align*}
-|\veck|^2 \hat{U}(t,\veck) = - \hat{\rho}(t,\veck)\,,\quad\Rightarrow\quad
\hat{\vecE} = -i\veck \hat{U} = -i\frac{\veck}{|\veck|^2} \hat{\rho}(t,\veck) \,,
\end{align*}
we rewrite~\eqref{eqn: LE system} into:
\begin{align}
\left(\veck\cdot \left(I+L[\hat{\mu}\tilde{\maxR}]\right) \veck \right) L[\hat{U}] & = L[\hat{S}] \,,
\end{align}
hence proving~\eqref{eqn:solvability}. \revise{The stability condition~\eqref{eqn: penrose cond} comes from inverting the Laplace transform~\cite{Davies_1978_transform}}.
\end{proof}

\subsection{Two case studies} \label{subsec:numerical_verify_penrose}
In this subsection, we compute~\eqref{eq:penrose_Pk_lam} for two equilibria, the Gaussian equilibrium and a ring-shaped equilibrium that generates Dory-Guest-Harris instability. Moreover, we restrict ourselves to the Bernstein modes, the Fourier modes of $\veck = (k_1, k_2,0)^\top$ that are transverse to the magnetic field. Consequently, the derivation of the rest of the paper is conducted in $2D2V$.

\begin{remark}
For another specific mode in the form of $(0,0,k_3)^\top$, our linear analysis recovers the previous results of Vlasov-Poisson~\cite{EINKEMMER-Li-Mouhot-Yue-2025}, where $\mu_{\parallel}(v_3)$ is two-stream instability or bump-on-tail instability along the magnetic field direction.
\end{remark} 

\subsubsection{Gaussian equilibrium.} We first consider Gaussian equilibrium of~\eqref{eqn: linearized VP}. The Gaussian equilibrium has been proved to have non-decaying electric energy for Bernstein modes~\cite{Bedrossian-Wang-2020-VP}. That is, for each $\veck=(k_1, k_2)^\top\in \mathbb{Z}^2$, the poles of $P(\veck, \lambda)$ are located along the imaginary line.

Let $\mu$ be any isotropic Gaussian distribution on $(v_1, v_2)^\top$, i.e.
\begin{align}\label{eq:iso Gaussian-theta1}
\mu(\vecv) = \dfrac{1}{2\pi\theta} e^{-\frac{v_1^2 + v_2^2}{2\theta}}\,. 
\end{align}
According to~\eqref{eq:penrose_Pk_lam},
\begin{align*}
P(\veck,\lambda)  
= & |\veck|^2 + |\veck|^2 \frac{1}{B_0}\int_0^{\infty} e^{-\lambda t} \hat{\mu}(\tilde{\maxR}(t) \veck) \sin(B_0 t) \,dt \,,\\
= & |\veck|^2 \left(1 +  \frac{1}{B_0}\int_0^{\infty} e^{-\lambda t} \exp{\left(-\frac{\theta}{B_0^2}(1-\cos(B_0 t))|\veck|^2\right)} \sin(B_0 t) \,dt \right)\,,
\end{align*}
where we used:
\begin{align*}
\veck \cdot \tilde{\maxR}(t)\veck & = 
\begin{pmatrix} k_1 \\ k_2 \end{pmatrix}^\top
\frac{1}{B_0}\begin{pmatrix}
\sin(B_0 t) & \cos(B_0 t)-1 \\ 
-\cos(B_0 t)+1 & \sin(B_0 t) 
\end{pmatrix} \begin{pmatrix} k_1 \\ k_2 \end{pmatrix} = \frac{\sin(B_0 t)}{B_0} |\veck|^2\,.
\end{align*}

\begin{figure}[htbp]
    \centering
     \begin{subfigure}[b]{0.4\textwidth}
         \centering
         \includegraphics[width=\textwidth]{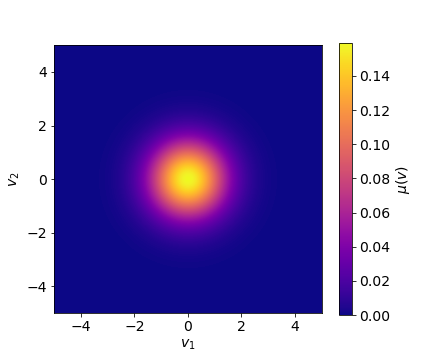}
    \caption{ }
    \label{fig: iso gaussian 2d plot}
     \end{subfigure}
     \hfill
     \begin{subfigure}[b]{0.4\textwidth}
         \centering
         \includegraphics[width=\textwidth]{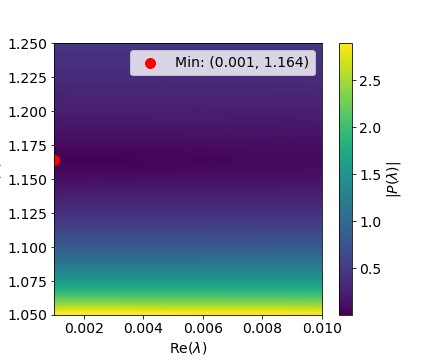}
    \caption{ }
    \label{fig: iso gaussian penrose}
    \end{subfigure}
    \caption{Isotropic Gaussian $\mu$ and $|P(\veck,\lambda)|$. The minimum is taken at $(\mathrm{Re}(\lambda) = 0.001, \mathrm{Im}(\lambda) = 1.164)$ with value $\min |P(0.001, 1.164)|=2\times 10^{-3}.$ }
    \label{fig: iso Gaussian}
\end{figure}

In the following we set $\theta=1$, $B_0=1$ and $\veck_{\perp}=(1,0)^\top$ without loss of generality, then
\begin{align*}
P(\veck,\lambda) & = 1 + \int_0^{\infty} e^{-\lambda t} e^{-1+\cos(t)} \sin(t) \,dt\,,
\end{align*}
In \revise{Figure}~\ref{fig: iso Gaussian}, we plot the equilibrium~\eqref{eq:iso Gaussian-theta1} and the evaluation of $P(\veck,\lambda)$ for $\lambda=\mathrm{Re}(\lambda)+i\mathrm{Im}(\lambda)$ with $(\mathrm{Re}(\lambda),\mathrm{Im}(\lambda))\in[0.001,0.1]\times[1.05, 1.25]$. Since poles for $P$ are complex conjugate pairs, we only plot them on the $\mathrm{Im}(\lambda)>0$. The plot shows that we can numerically find a root at $(\mathrm{Re}\lambda = 0.001, \mathrm{Im}(\lambda)=1.164)$. This is consistent with the analytical prediction in~\cite{Bedrossian-Wang-2020-VP} that suggests all roots should be along the imaginary line\footnote{Numerically, we cannot compute $P(\veck,\lambda)$ for $\mathrm{Re}(\lambda)=0$.}.

\subsubsection{Dory-Guest-Harris instability \label{subsec:DGH-instability-penrose}}
We also demonstrate the computation of~\eqref{eq:penrose_Pk_lam} for the ring-shaped equilibrium as discussed in~\cite{DGH-1965-PRL}, 
\begin{equation}\label{eqn: DGH-fv}
\mu(\vecv) = \frac{1}{\pi \alpha_{\perp}^2 j!} \left(\frac{v_1^2 + v_2^2}{\alpha_{\perp}^2}\right)^j \exp{\left(-\frac{v_1^2 + v_2^2}{\alpha_{\perp}^2}\right)}\,.
\end{equation}
A special case for this equilibrium is when $j=0$, when it recovers the Gaussian equilibrium. It was further identified~\cite{DGH-1965-PRL} the equilibrium~\eqref{eqn: DGH-fv} continue being unconditionally stable for $j\in\{0,1,2\}$. However, for $j\geq 3$, the equilibrium~\eqref{eqn: DGH-fv} can be unstable, and this instability is termed the Dory-Guest-Harris instability. Intuitively, for larger $j$, the distribution becomes more concentrated on a ring~\cite{Tataronis-Crawford-1965-abs-instable, Tataronis_Crawford_1970}, and this concentration leads to instability. The cases of $j=3,4,5,6$ were thoroughly studied in~\cite{VOGMAN2014101}.

We leave detailed derivation to Appendix~\ref{appx:DGH_Fourier_compute}. In the following example we set $j=6, \alpha_{\perp}=\sqrt{1/3}, B_0=0.05$, then 
\begin{equation}\label{eq:DGH6_equilibrium}
    \mu(\vecv) = \frac{3}{\pi 6!} \left(\frac{v_1^2 + v_2^2}{1/3}\right)^6 \exp{\left(-\frac{v_1^2 + v_2^2}{1/3}\right)}\,.
\end{equation}
In this case, according to the derivation in Appendix~\ref{appx:DGH_Fourier_compute}, we have
\begin{align*}
P(\veck, \lambda)&= |\veck|^2\left(1 + \int_0^{\infty} \exp{\left(-\lambda t -z\right)} M(-6,1, z) \frac{\sin(B_0 t)}{B_0}\,dt\right)\,.
\end{align*}
where $z = \frac{1}{6 B_0^2}|\veck|^2(1-\cos(B_0 t))$.

\begin{figure}[htbp]
    \centering
    \begin{subfigure}[b]{0.38\textwidth}
        \centering
        \includegraphics[width=\textwidth]{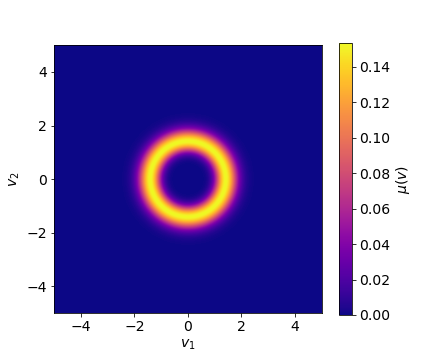}
        \caption{ }
    \end{subfigure}
    \hfill
    \begin{subfigure}[b]{0.38\textwidth}
        \centering
        \includegraphics[width=\textwidth]{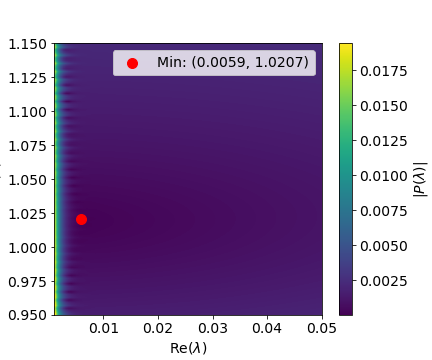}
        \caption{ }
    \end{subfigure}
    \caption{Equilibrium distribution~\eqref{eq:DGH6_equilibrium} and $|P(\veck,\lambda)|$~\eqref{eq:penrose_Pk_lam}. The minimum is taken at $(\lambda_R=0.0059, \lambda_I=1.0207)$ with minimal value $1.5 \times 10^{-5}.$}
    \label{fig:DGH_dist_disp_ksq1}
\end{figure}

Again without loss of generality, we fix $|\veck|^2=1$, i.e. $\veck=(1,0)^\top$ or $\veck=(0,1)^\top$, and numerically compute $P$ as a function of $\lambda$~\eqref{eqn: penrose cond}. In Figure~\ref{fig:DGH_dist_disp_ksq1} we plot both the distribution function~\eqref{eq:DGH6_equilibrium} and $P(\veck=(1,0),\lambda)$. Clearly, in this case, we find a root with a positive real component, indicating the instability of this equilibrium.

\section{Control with an external electric field \label{sec:control_external_E}} 
 
Our strategy to suppress instability in the magnetized Vlasov-Poisson system is to add an external electric field. 
Add an external electric field $-\nabla_{\vecx} \Phi(t,\vecx)$, generated by some electric potential $\Phi$, to the system~\eqref{eqn: linearized VP}, the system is:
\begin{align}
\partial_t f + \vecv\cdot \nabla_{\vecx} f + (v_2 B_0 \partial_{v_1} f - v_1 B_0\partial_{v_2} f) + (\vecE -\nabla_{\vecx} \Phi)\cdot \nabla_{\vecv} \mu(\vecv) & = 0\,, \label{eqn: linearized VP w/ H}\\
\nabla_{\vecx} \cdot \vecE & = \int_{\R^3} f\,d\vecv\,.
\end{align}
Following the same linear analysis as in Proposition~\ref{prop:penrose_cond}, we have:
\begin{align}\label{eq:U_dispersion_k_lam_Phi}
|\veck|^2 L[\hat{U}] - L[\hat{S}] + (L[\hat{U}] + L[\hat{\Phi}]) \veck \cdot L[\hat{\mu}\tilde{R}] \veck & = 0\,,
\end{align}
where we used $\hat{\vecE}=-i\veck\hat{U}$. Notice $1 = \frac{\veck\cdot \veck}{|\veck|^2}$ and recall the definition of our dispersion function~\eqref{eq:penrose_Pk_lam}, the equation becomes:
\begin{align}\label{eq:penrose_source_Pk}
P(\veck,\lambda) L[\hat{U}] &= L[\hat{S}] - L[\hat{\Phi}] \veck\cdot L[\hat{\mu}\tilde{R}] \veck \,. 
\end{align}
Compare this equation with~\eqref{eqn:solvability}, we quickly notice that the only difference is that the right-hand side is modified with a new source term that depends on the external potential $\Phi$. Physically, one could wish to have the solution to be close to the free-streaming, which requires $U$, the self-generated electric field, to satisfy $\Delta U=S$. For computational convenience, it would be easier to translate the equation into:
\begin{align}\label{eq:penrose_source_ksq}
P(\veck,\lambda)\left(|\veck|^2 L[\hat{U}]-L[\hat{S}]\right) 
&= - \left(|\veck|^2 L[\hat{\Phi}] + L[\hat{S}]\right) \veck\cdot L[\hat{\mu} \tilde{R}] \veck\,,
\end{align}
Recall Proposition~\ref{prop:penrose_cond}, instability is observed when $P$ has roots on $\mathrm{Re}(\lambda)>0$ side of the plane. One straightforward strategy to eliminate this instability is to set $\Phi$ so to have the entire source term in~\eqref{eq:penrose_source_ksq} cancels out the roots of $P$. Mathematically, this is to set:
\begin{equation}\label{eq:control_lim_c}
\lim_{\lambda\to\lambda_0}\frac{\left(|\veck|^2 L[\hat{\Phi}] + L[\hat{S}]\right) \veck\cdot L[\hat{\mu} \tilde{R}] \veck}{P(\veck,\lambda)} = c\,,\quad\forall\veck, \lambda_0\quad{s.t.}\quad P(\veck,\lambda_0)=0\,.
\end{equation}

The constant $c$ here can be arbitrary. Since we do not have control over $\veck\cdot L[\hat{\mu} \tilde{R}] \veck$, the control strategy becomes manipulating $|\veck|^2 L[\hat{\Phi}] + L[\hat{S}]$ only. In the subsequent two sections, we propose two simplest options to achieve~\eqref{eq:control_lim_c}.



\subsection{Control Strategy 1: set 
\texorpdfstring{\(c\equiv 0\)}{0} in \texorpdfstring{\eqref{eq:control_lim_c}}{(3.6)}.
}  \label{sec:control_c_0}
This, according to~\eqref{eq:control_lim_c}, leads to:
\begin{align}\label{eq:control_Phi_c0}
|\veck|^2 L[\hat{\Phi}] + L[\hat{S}] = 0\,,\quad \forall \veck \in \mathbb{Z}^3, \lambda\,.
\end{align}
Take the inverse Laplace transform and the inverse Fourier transform, and this requirement becomes
\[
-\Delta \Phi = S\,,
\]
where $S$ is defined as in~\eqref{eq:S_source_free_sol}. When~\eqref{eq:control_Phi_c0} holds, the right hand side of~\eqref{eq:penrose_source_ksq} is zero, and we will also obtain:
\[
\Delta U = S\,.
\]
This means the self-generated electric field is completely zeroed out by the external field $\nabla_{\vecx}\Phi$. Plugging them in~\eqref{eqn: linearized VP w/ H}, we find the solution recovers the free-streaming solution, as described in the following lemma.

\begin{lemma}\label{lem:recovery}
The controlled system~\eqref{eqn: linearized VP w/ H} with $\Phi$ designed in~\eqref{eq:control_Phi_c0} coincides with the free-streaming solution~\eqref{eq:free_stream_eq} for all time.
\end{lemma}
\begin{proof}
Define the difference of solution $f_\delta = f - f_{\mathrm{f}}$, and we obtain the system of $f_\delta$ by subtracting~\eqref{eqn: linearized VP w/ H} from~\eqref{eq:free_stream_eq}:
\begin{align}
\partial_t f_\delta + \vecv\cdot\nabla_{\vecx} f_{\delta} + (\vecv\times \magB)\cdot \nabla_{\vecv} f_{\delta} + (\vecE[f] -\nabla_{\vecx}\Phi) \cdot \nabla_{\vecv} \mu & = 0\,,\\
f_{\delta} (t=0, \vecx,\vecv) & = 0\,.
\end{align}
Using the linearity of $\vecE$, we compute 
\[\vecE[f] -\nabla_{\vecx} \Phi = \vecE[f] - \vecE[f_{\mathrm{f}}]= \vecE[f_\delta](t,\vecx)\,.\]
The system is thus linear in $f_\delta$. Since the initial data is $0$, the solution is always zero: $f_\delta\equiv 0$.
\end{proof}

\begin{remark}\label{rmk:nonlinear_stream}
    We should note that this control strategy is also effective for the nonlinear Vlasov-Poisson system. Indeed, by revising~\eqref{eqn: linearized VP w/ H} into:
    \[
    \partial_t f + \vecv\cdot \nabla_{\vecx} f + (v_2 B_0 \partial_{v_1} f - v_1 B_0\partial_{v_2} f) + (\vecE -\nabla_{\vecx} \Phi)\cdot \nabla_{\vecv} f = 0\,.
    \]
    Setting $\Phi=U$, we still recover the free-streaming solution for all time.
\end{remark}

By linking the controlled system~\eqref{eqn: linearized VP w/ H} with the free-streaming solution, we can explicitly analyze the in-time dynamics of the electric energy.

\begin{theorem}\label{thm:free_stream_control_energy}
Let $f$ be the solution to~\eqref{eqn: linearized VP w/ H} with the external field prepared in~\eqref{eq:control_Phi_c0}, equipped with separable initial profile: $f_0(\vecx,\vecv) = \chi(\vecx)\mu(\vecv)$ where $\mu(\vecv)$ is Gaussian with temperature $\theta$, and $\chi(\vecx)$ is periodic in space and $\hat{\chi}$ is bounded. Denote $\mathcal{E}$ as the electric energy defined in~\eqref{eqn:electric_energy}. Let $\mathcal{E}_{\veck}(t):=\frac{1}{2}|\hat{\vecE}(t,\veck)|^2$ be the $\veck=(k_1,k_2,k_3)$-th mode of $\mathcal{E}$. Then:
\begin{itemize}
    \item For non-Bernstein modes, i.e. $k_3\not=0$, electric energy $\mathcal{E}_{\veck}(t)$ exponentially decays in time.
    \item For Bernstein modes, i.e. $k_3=0$, electric energy $\mathcal{E}_{\veck}(t)$ presents a periodic pattern in time with the period being $\frac{2\pi}{B_0}$, the inverse of the standard gyrofrequency in plasma physics. 
\end{itemize}
\end{theorem}

This theorem suggests, in the long time, the electric energy is predominantly governed by modes perpendicular to $\magB$.

\begin{proof}
According to Lemma~\ref{lem:recovery}, the controlled solution recovers the free-streaming solution, and thus, recall~\eqref{eq:free_stream_sol}:
\[
f(t,\vecx,\vecv)=f_{\mathrm{f}}(t,\vecx,\vecv) = \chi(\vecx_\perp-(\maxM(t)+\maxQ) \maxR^{-1}(B_0 t)\vecv_\perp, x_3-v_3t)\mu(\maxR^{-1}(B_0 t)\vecv_\perp, v_3)\,.
\]
Take the Fourier transform and integrate in the velocity space:
\begin{align*}
\hat{\rho}(t,\veck) & = \int e^{-i\veck\cdot \vecx} \int f(t,\vecx,\vecv)\,d\vecv\,d\vecx\,,\\
& = \int e^{-i\veck\cdot \vecx} \int \chi(\vecx_\perp-(\maxM(t)+\maxQ)R^{-1}(B_0 t)\vecv_\perp, x_3-v_3t)\mu(\maxR^{-1}(B_0 t)\vecv_\perp, v_3) \,d\vecv\,d\vecx\,,\\
& = \hat{\chi}(\veck) \hat{\mu}((\maxM(t)+\maxQ)^\top \veck_\perp, k_3 t)\,,
\end{align*}
where $\hat{\chi}$ is bounded and the Fourier transform of $\mu$ can be explicitly computed:
\begin{align*}
\hat{\mu}((\maxM(t) + \maxQ)^\top \veck_\perp, k_3 t) = \exp{\left(-\theta(1-\cos(B_0t))(k_1^2+k_2^2) - \frac{\theta}{2}k_3^2 t^2\right)}\,.
\end{align*}
\begin{itemize}
\item For non-Bernstein modes, the above is upper bounded,
\[\hat{\mu}((\maxM(t)+\maxQ)^\top \veck_\perp, k_3 t) \leq e^{- \frac{\theta}{2}k_3^2 t^2}\,,\]
hence
\begin{align*}
\mathcal{E}_{\veck}(t) = \frac{1}{2}|\vecE_{\veck}|^2 = \frac{1}{2|\veck|^2}\hat{\rho}^2(t,\veck) \leq \frac{1}{2}|\hat{\chi}(\veck)|^2 |\hat{\mu}((\maxM(t)+\maxQ)^\top \veck_\perp, k_3 t)|^2 \leq C e^{-A t^2}\,,
\end{align*}
for some constant $A, C>0$.
\item For Bernstein modes, i.e. $k_3 = 0$,
\[\hat{\mu}((\maxM(t)+\maxQ)^\top \veck_\perp, k_3 t) = \exp{\left(-\theta(1-\cos(B_0t))(k_1^2+k_2^2)\right)} \leq 1\,,\]
then \revise{the electric energy is periodic and bounded:}
\begin{align*}
\mathcal{E}_{\veck}(t) & 
= \frac{1}{2 |\veck_{\perp}|^2}|\hat{\chi}(\veck)|^2 \exp{\left(-2\theta(1-\cos(B_0t))(k_1^2+k_2^2)\right)}
\leq \frac{1}{2}|\hat{\chi}(\veck)|^2 \,.
\end{align*}
\revise{The periodicity is shown in the $\cos$ function in the exponential factor, with the period being $\frac{2\pi}{B_0}$.}
\end{itemize}
\end{proof}

\subsection{Control Strategy 2: set 
\texorpdfstring{$c$}{c} 
to be constant in \texorpdfstring{\eqref{eq:control_lim_c}}{(3.6)}.}
\label{subsec:control_strategy_2c}

\revise{To provide an additional concrete realization of the general control design, we propose this section control strategy to set $c$ to be a constant in~\eqref{eq:control_lim_c}. Allowing $c$ to be an arbitrary constant gives a greater flexibility for exercising control.}

According to~\eqref{eq:control_lim_c}, this is to set:
\begin{align}\label{eq:control2_c_formula}
|\veck|^2 L[\hat{\Phi}] + L[\hat{S}] & = c \veck\cdot (I+L[\hat{\mu}\tilde{\maxR}]) \veck\,.
\end{align}
We apply inverse Laplace transform, and notice $L^{-1}[1]=\delta(t)$:
\begin{align*}
\hat{\Phi}(t,\veck) & = - \frac{1}{|\veck|^2}\hat{f}_0(\veck,(\maxM(t)+\maxQ)^\top \veck_\perp, k_3) 
+ c \delta(t) 
+ c \frac{1}{|\veck|^2} \hat{\mu}(\tilde{\maxR}(t)\veck) \veck\cdot \tilde{\maxR}(t) \veck \,,
\end{align*}
Using this choice of $\Phi$, we can compute the self-generated field. This is done by plugging it in~\eqref{eq:penrose_source_ksq} to see:
\begin{equation*}
\begin{aligned}
|\veck|^2 L[\hat{U}] - L[\hat{S}] = - c\veck\cdot L[\hat{\mu}\tilde{\maxR}]\veck\,.
\end{aligned}
\end{equation*}
Therefore, the self-generated potential has the free-streaming component and a component from propagating the equilibrium state:
\begin{equation}\label{eq:control_2_Uhat}
\hat{U} = \frac{\hat{S}}{|\veck|^2} - \frac{1}{|\veck|^2} c \veck\cdot \hat{\mu}\tilde{\maxR}\veck\,.
\end{equation}
\begin{remark}
We should stress that since $\tilde{\maxR}$ is a $3\times 3$ matrix, the $\veck\cdot\tilde{\maxR}\veck$ term cannot easily pull out the $|\veck|^2$ component. With more calculation, one can derive:
\begin{equation}\label{eq:kRk}
\veck\cdot\tilde{\maxR}\veck = \frac{\sin(B_0 t)}{B_0}|\veck_{\perp}|^2 + k_3^2 t\,.
\end{equation}
This structure prevents us from directly taking the Fourier inverse transform for an explicit solution for $U$, as was done in~\cite{EINKEMMER-Li-Mouhot-Yue-2025} for the Vlasov-Poisson system without external $\magB$.
\end{remark}


\revise{The performance of this control is not as clean as that of Control Strategy 1. Since it is not able to recover the free-streaming solution, this controlled system maintains nonlinear features, and the performance needs to be discussed case-by-case. Recall the definition of the electric energy of the $\veck$-th mode ($\mathcal{E}_{\veck}(t)
= \frac{1}{2}|\veck|^2|\hat{U}_{\veck}|^2$) and the Poisson equation~\eqref{eqn: linearized E}, following~\eqref{eq:control_2_Uhat}, we have:
\begin{equation*}
\begin{aligned}
\mathcal{E}_{\veck}(t) & 
= \frac{1}{2}|\veck|^2|\hat{U}_{\veck}|^2 
= \frac{1}{2}\left|\hat{S} - c \veck\cdot \hat{\mu}\tilde{\maxR}\veck\right|^2 
\leq |\hat{S}|^2 + c^2 \left| \veck\cdot \hat{\mu}\tilde{\maxR}\veck\right|^2\\
& \leq |\hat{f}_0(\veck,(\maxM(t) + \maxQ)^\top \veck_{\perp}, k_3)|^2 + c^2 |\hat{\mu}|^2 \left| \frac{1}{B_0}|\veck_{\perp}|^2 + k_3^2 t\right|^2\,.
\end{aligned}
\end{equation*}
where we used the definition of $S$, \eqref{eq:free_stream_sol} and~\eqref{eq:kRk}.  If the initial profile is assumed to be of a factored form as in Theorem~\ref{thm:free_stream_control_energy} with $\mu$ being a Gaussian, we obtain the following estimate of the temporal behavior of the electric energy:
\begin{equation*}
\begin{aligned}
\mathcal{E}_{\veck}(t) &\leq \exp{\left(-2\theta(1-\cos(B_0t))(k_1^2+k_2^2) - \frac{\theta}{2}k_3^2 t^2\right)}\left(|\chi(\veck)|^2 + c^2\left| \frac{1}{B_0}|\veck_{\perp}|^2 + k_3^2 t\right|^2\right)\,.
\end{aligned}
\end{equation*}
That is, for non-Bernstein modes, $k_3\not=0$, the electric energy of the system can be controlled to decay sub-exponentially, at the rate of $O(t^2 e^{-k_3^2 t^2})$. For Bernstein modes, $k_3=0$, the electric energy is again controlled to be bounded, $\mathcal{E}_{\veck}(t) \leq |\chi(\veck)|^2 + \frac{c^2}{B_0^2}|\veck_{\perp}|^4$, but no-decay can be said.
}

\section{Numerical experiments \label{sec:numerics}}
In this section, we present numerical evidence of the two proposed control strategies.

We choose numerical examples on a $2D2V$ domain by simulating the system for $\vecx = (x_1, x_2)^\top\in\mathbb{T}^2, \vecv=(v_1, v_2)^\top\in\mathbb{R}^2$: 
\[
\partial_t F + \vecv\cdot \nabla_{\vecx} F + (v_2 B_0 \partial_{v_1} F - v_1 B_0\partial_{v_2} F) + \vecE\cdot \nabla_{\vecv} F(\vecv) = 0\,.
\]
We test our solution using both Gaussian equilibrium~\eqref{eq:iso Gaussian-theta1} and the ring-shaped equilibrium~\eqref{eq:DGH6_equilibrium} for $\mu(\vecv)$. We also set the initial condition to be a small perturbation from such equilibrium states:
\begin{align}\label{eq:ic_perturb_h}
F(t=0, \vecx, \vecv) = (1+ h(\vecx))\mu(\vecv) \,,
\end{align}
where $h(\vecx)$ denotes the initial perturbation in space, and $|h(\vecx)|$ is taken to be small.

The nonlinear system is computed using a semi-Lagrangian scheme with Strang splitting~\cite{CROUSEILLES20101927, EINKEMMER2019937} for the time integration. This means from one step to the next, we first solve
\begin{align} \label{eqn: Hf update}
\partial_t F + \vecv\cdot\nabla_{\vecx} F = 0\,,
\end{align}
by $\Delta t/2$, and it is followed by solving
\begin{align} \label{eqn: HB update}
\partial_t F + (\vecv\times \magB)\cdot\nabla_{\vecv} F = 0\,,
\end{align}
by $\Delta t/2$ and 
\begin{align} \label{eqn: HE update}
\begin{cases}
    \vecE = - \nabla_{\vecx} \Delta_{\vecx}^{-1} \left(1-\int F \,d\vecv\right)\,,\\
\partial_t F + \vecE \cdot\nabla_{\vecv} F = 0\,,
\end{cases}
\end{align}
by $\Delta t$, before solving~\eqref{eqn: HB update} and~\eqref{eqn: Hf update} each by $\Delta t/2$. In $\vecx$, we use the Fourier spectrum method, and in $\vecv$, we perform linear interpolation. 

The code is written in JAX, and the computation was performed on an NVIDIA Quadro RTX 6000 GPU with 24 GB of memory. \revise{The data and code for results presented here are openly available in the GitHub repository\footnote{\url{https://github.com/Peiyi-wisc/Magnetized_VP_control.git}}, and archived in Zenodo\footnote{\url{https://doi.org/10.5281/zenodo.15109942}}.}

\subsection{Dory-Guest-Harris instability.}
This section is dedicated to the cases where $\mu$ is chosen as a ring-shaped equilibrium~\eqref{eq:DGH6_equilibrium} with index $j=6$ (hence termed DGH6). The two subsections~\ref{sec:DGH_sinx_control} and~\ref{sec:DGH_gaussx_control} use different initial perturbations, and we apply two different control strategies to showcase the prediction of our theory.

\subsubsection{Initial perturbation as a sine wave in space\label{sec:DGH_sinx_control}}
In this example we set $B_{0}= 0.05$, and use the following initial condition:
\begin{equation}\label{eq:nonlinear_ic_k01_a00001}
F_0(\vecx, \vecv) = (1 + 0.001 \sin(0.1 x_1)) \frac{1}{\pi 6!/ 3} \left(\frac{v_1^2 + v_2^2}{1/3}\right)^6 \exp{\left(-\frac{v_1^2 + v_2^2}{1/3}\right)}\,.
\end{equation}
For simulation we use $x_1, x_2 \in [-10\pi, 10\pi], v_1, v_2\in[-5,5], N_{x_1}=N_{x_2}=64, N_{v_1}=N_{v_2}=64$.

\begin{figure}[htbp]
    \centering
    \begin{subfigure}[b]{0.42\textwidth}
        \centering
        \includegraphics[width=\textwidth]{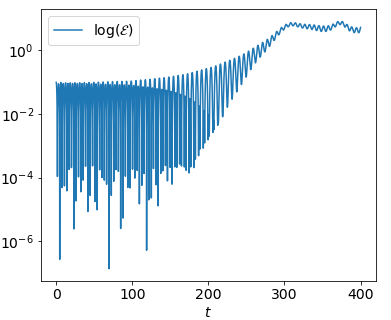}
        \caption{ }
    \end{subfigure}
    \hfill
    \begin{subfigure}[b]{0.42\textwidth}
        \centering
        \includegraphics[width=\textwidth]{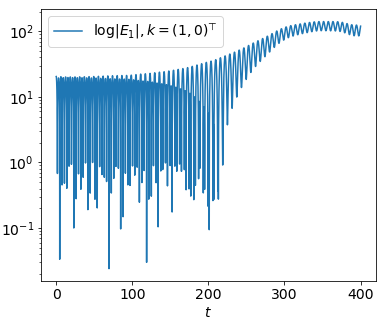}
        \caption{ }
    \end{subfigure}
    \caption{The left panel shows the evolution of the electric energy $\mathcal{E}(t)$, and the right panel shows the first coordinate of $|\vecE_{(1,0)^\top}(t)|$. Both plots are presented on the semi-log scale.}
    \label{fig:DGH_energy_no_control}
\end{figure}
\begin{figure}[htbp]
    \centering
    \begin{subfigure}[b]{0.32\textwidth}
        \centering
        \includegraphics[width=\textwidth]{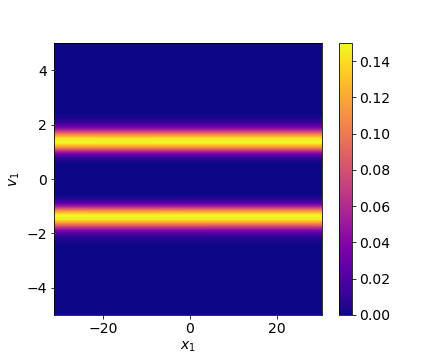}
        \caption{$t=0$}
    \end{subfigure}
    \hfill
    \begin{subfigure}[b]{0.32\textwidth}
        \centering
        \includegraphics[width=\textwidth]{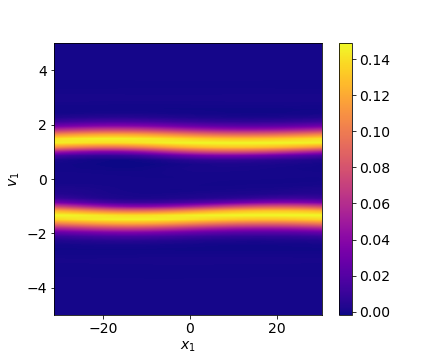}
        \caption{$t=200$}
    \end{subfigure}
    \hfill
    \begin{subfigure}[b]{0.32\textwidth}
        \centering
        \includegraphics[width=\textwidth]{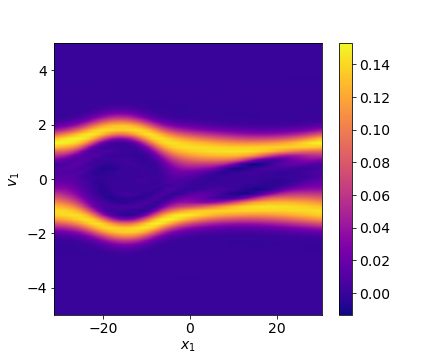}
        \caption{$t=300$}
    \end{subfigure}
    \\
    \begin{subfigure}[b]{0.32\textwidth}
        \centering
        \includegraphics[width=\textwidth]{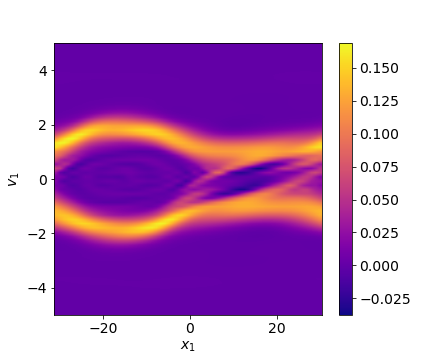}
        \caption{$t=350$}
    \end{subfigure}
    \hfill
    \begin{subfigure}[b]{0.32\textwidth}
        \centering
        \includegraphics[width=\textwidth]{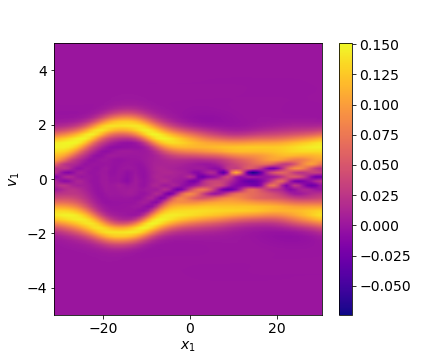}
        \caption{$t=375$}
    \end{subfigure}
    \hfill
    \begin{subfigure}[b]{0.32\textwidth}
        \centering
        \includegraphics[width=\textwidth]{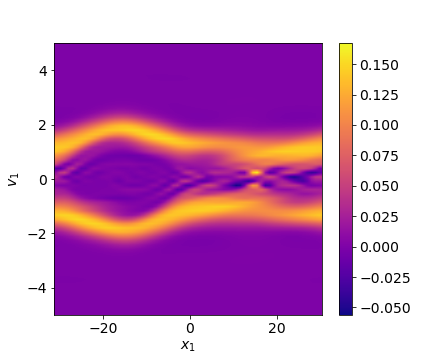}
        \caption{$t=400$}
    \end{subfigure}
    \caption{Distribution $f(x_1, x_2=0, v_1, v_2=0)$ snapshot at different time steps, without control. The distribution is asymmetric in the spatial domain, and the effect is rooted in the asymmetry in the perturbation in the initial data.}
    \label{fig:DGH6_fxvx_no_control}
\end{figure}
\begin{figure}[htbp]
    \centering
    \begin{subfigure}[b]{0.46\textwidth}
        \centering
        \includegraphics[width=\textwidth]{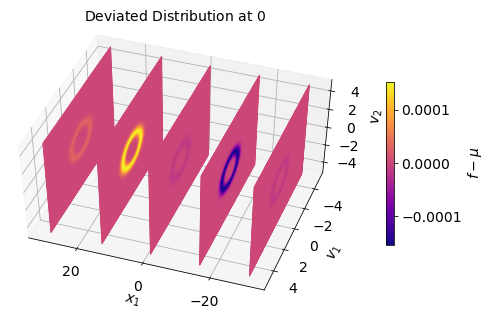}
        \caption{$t=0$}
    \end{subfigure}
    \hfill
    \begin{subfigure}[b]{0.46\textwidth}
        \centering
        \includegraphics[width=\textwidth]{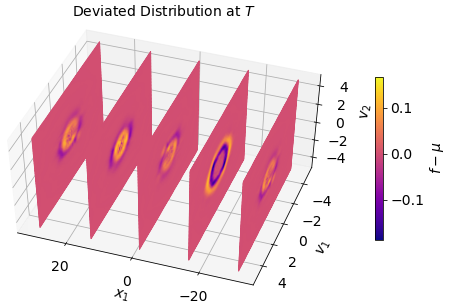}
        \caption{$t=400$}
    \end{subfigure}
    \caption{Distribution $f(x_1, x_2=0, v_1, v_2)$ at five $x_1$ locations, without control.}
    \label{fig:DGH6_f_dist_no_control}
\end{figure}

Without control, we observe that the electric energy exponentially grows in time, as shown in Figure~\ref{fig:DGH_energy_no_control}. In Figure~\ref{fig:DGH6_fxvx_no_control}, we present a slice of distribution $f(x_1, v_1)$ by setting $(x_2,v_2)=(0,0)$. The distribution starts showing turbulent features as early as $t=300$, and deforms as time evolves. In Figure~\ref{fig:DGH6_f_dist_no_control}, we plot a 3D plot by fixing $x_2=0$, and it is clear that the ring-shape distribution is deformed at $T=400$.

\begin{figure}[htbp]
    \centering
    \begin{subfigure}[b]{0.4\textwidth}
        \centering
        \includegraphics[width=\textwidth]{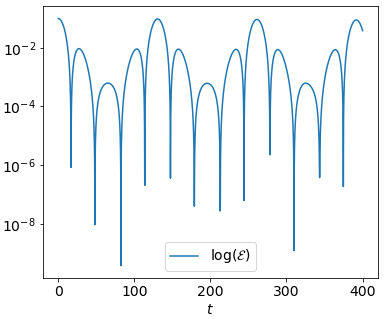}
        \caption{ }
    \end{subfigure}
    \hfill
    \begin{subfigure}[b]{0.4\textwidth}
        \centering
        \includegraphics[width=\textwidth]{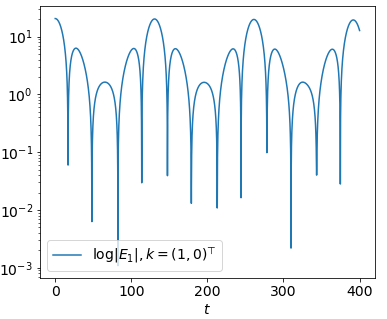}
        \caption{ }
    \end{subfigure}
    \caption{Evolution of the electric energy $\mathcal{E}(t)$, and the first coordinate of $|\vecE_{(1,0)^\top}(t)|$ in semi-log scale, with Control Strategy 1 that $c=0$. }
    \label{fig:DGH6_energy_control1}
\end{figure}

\begin{figure}[htbp]
    \centering
    \begin{subfigure}[b]{0.32\textwidth}
        \centering
        \includegraphics[width=\textwidth]{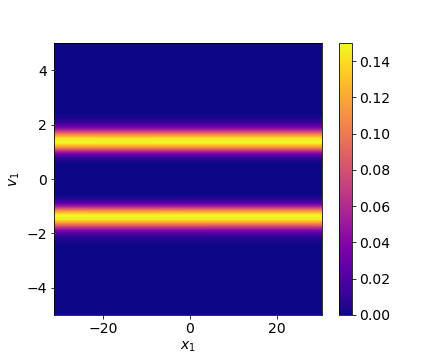}
        \caption{$t=0$}
    \end{subfigure}
    \hfill
    \begin{subfigure}[b]{0.32\textwidth}
        \centering
        \includegraphics[width=\textwidth]{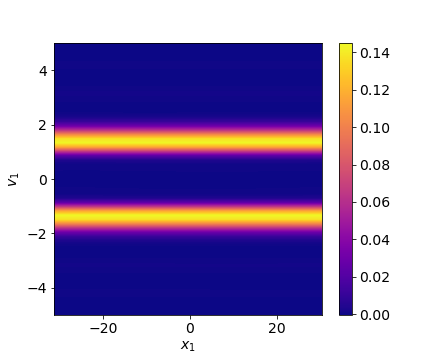}
        \caption{$t=300$}
    \end{subfigure}
    \hfill
    \begin{subfigure}[b]{0.32\textwidth}
        \centering
        \includegraphics[width=\textwidth]{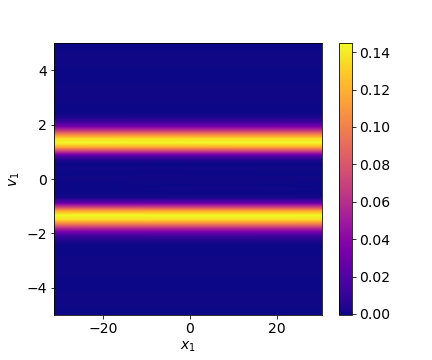}
        \caption{$t=400$}
    \end{subfigure}
    \caption{Distribution $f(x_1, x_2=0, v_1, v_2=0)$ snapshot at different time steps, with Control Strategy 1 that $c=0$.}
    \label{fig:DGH6_fxvx_control_param0}
\end{figure}

\begin{figure}[htbp]
    \centering
    \begin{subfigure}[b]{0.46\textwidth}
        \centering
        \includegraphics[width=\textwidth]{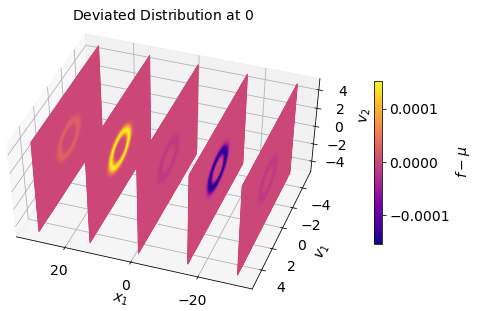}
        \caption{$t=0$}
    \end{subfigure}
    \hfill
    \begin{subfigure}[b]{0.46\textwidth}
        \centering
        \includegraphics[width=\textwidth]{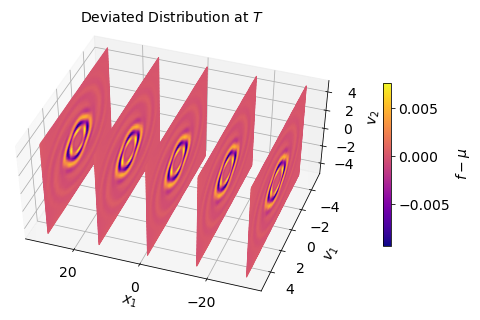}
        \caption{$t=400$}
    \end{subfigure}
    \caption{Distribution $f(x_1, x_2=0, v_1, v_2)$ at five $x_1$ locations, with control $c=0$.}
    \label{fig:DGH6_f_dist_control1}
\end{figure}

We apply Control Strategy 1 by setting $c=0$ first. This eliminates all the poles, and as suggested by Theorem~\ref{thm:free_stream_control_energy}, it should recover the free-streaming solution, whose energy is expected to be bounded and periodic (Bernstein modes). In Figures~\ref{fig:DGH6_energy_control1},~\ref{fig:DGH6_fxvx_control_param0} and~\ref{fig:DGH6_f_dist_control1}, we respectively show the counterparts for the case without the control. More specifically, it is clear in Figure~\ref{fig:DGH6_energy_control1} that energy is no longer growing. While the slice distribution of $f(x_1, x_2=0,v_1,v_2=0)$ continues showing the two-beam structure, the full ring is deformed, as seen in Figure~\ref{fig:DGH6_fxvx_control_param0} and Figure~\ref{fig:DGH6_f_dist_control1}.

\subsubsection{Initial perturbation as a Gaussian in space.}\label{sec:DGH_gaussx_control}
\begin{figure}[htbp]
    \centering
    \begin{subfigure}[b]{0.4\textwidth}
        \centering
        \includegraphics[width=\textwidth]{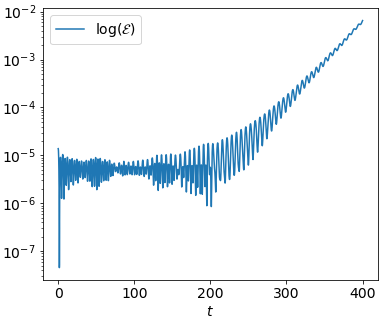}
        \caption{ }
    \end{subfigure}
    \hfill
    \begin{subfigure}[b]{0.4\textwidth}
        \centering
        \includegraphics[width=\textwidth]{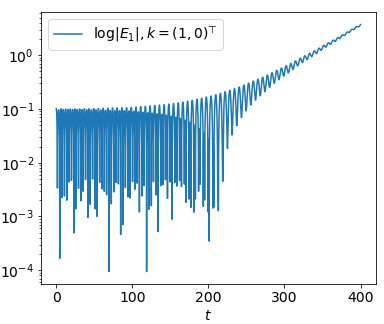}
        \caption{ }
    \end{subfigure}
    \caption{Evolution of the electric energy $\mathcal{E}(t)$, and the first coordinate of $|\vecE_{(1,0)^\top}(t)|$ in semi-log scale, without control.}
    \label{fig:DGH_gaussx_energy_no_control}
\end{figure}

\begin{figure}[htbp]
    \centering
    \begin{subfigure}[b]{0.32\textwidth}
        \centering
        \includegraphics[width=\textwidth]{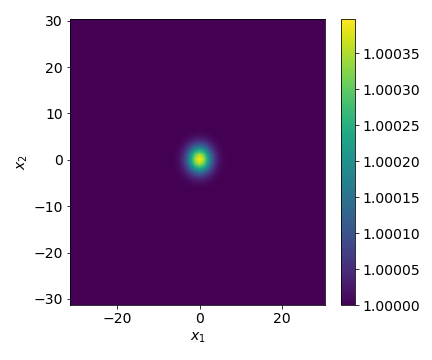}
        \caption{$t=0$}
    \end{subfigure}
    \hfill
    \begin{subfigure}[b]{0.32\textwidth}
        \centering
        \includegraphics[width=\textwidth]{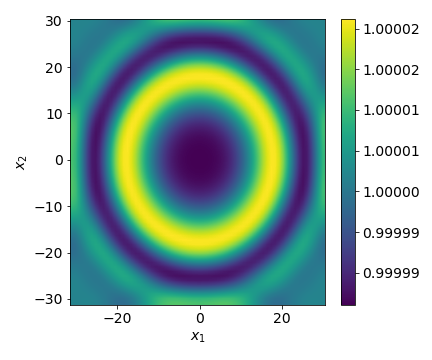}
        \caption{$t=10$}
    \end{subfigure}
    \hfill
    \begin{subfigure}[b]{0.32\textwidth}
        \centering
        \includegraphics[width=\textwidth]{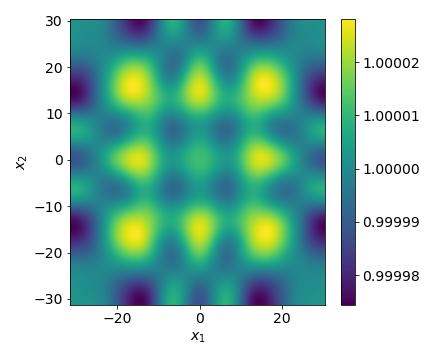}
        \caption{$t=50$}
    \end{subfigure}
    \\
    \begin{subfigure}[b]{0.32\textwidth}
        \centering
        \includegraphics[width=\textwidth]{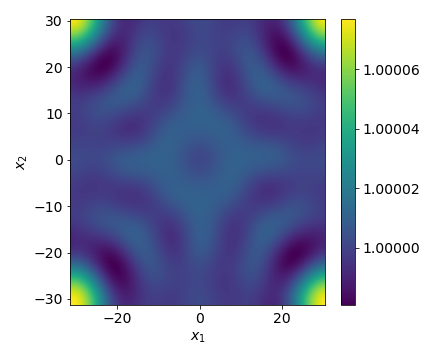}
        \caption{$t=200$}
    \end{subfigure}
    \hfill
    \begin{subfigure}[b]{0.32\textwidth}
        \centering
        \includegraphics[width=\textwidth]{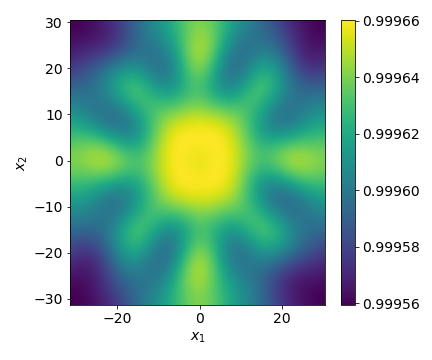}
        \caption{$t=300$}
    \end{subfigure}
    \hfill
    \begin{subfigure}[b]{0.32\textwidth}
        \centering
        \includegraphics[width=\textwidth]{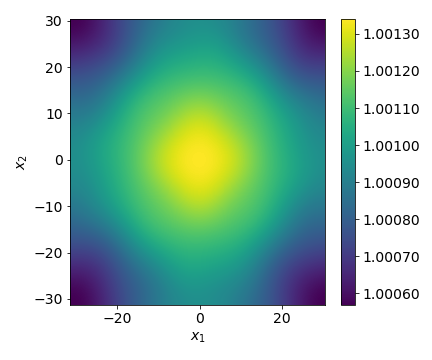}
        \caption{$t=400$}
    \end{subfigure}
    \caption{Density $\rho(x_1, x_2)$ snapshot at different time steps, without control.}
    \label{fig:DGH6_gaussx_nonlinear_rhoxy}
\end{figure}

We then examine an example with the ring-shaped equilibrium and perturbed initially by a Gaussian in space $\vecx$:
\[
F_0(\vecx, \vecv) = \left(1+0.01 \frac{1}{2\pi \theta_{0}}e^{-\frac{\vecx^2}{2 \theta_{0}}}\right)
\frac{1}{\pi 6!/ 3} \left(\frac{v_1^2 + v_2^2}{1/3}\right)^6 \exp{\left(-\frac{v_1^2 + v_2^2}{1/3}\right)}\,.
\]
We set $B_0=0.05$. From analysis in Section~\ref{subsec:DGH-instability-penrose}, this equilibrium is not stable, and we expect the dynamics to present the Dory-Guest-Harris instability, and that the electric energy $\mathcal{E}(t)$ and $|\vecE_1(\veck)|$ for $\veck=(1,0)^\top$ are expected to grow exponentially in time. This is shown in Figure~\ref{fig:DGH_gaussx_energy_no_control}. We also plot the evolution of density $\rho$ in Figure~\ref{fig:DGH6_gaussx_nonlinear_rhoxy}. The initial perturbation is a Gaussian at the origin, and the magnitude variation in $\rho$ highly concentrates at $\vecx=0$. Once we kick off the dynamics, the Gaussian beam starts propagating outward and forms a ring-shaped distribution. The magnitude of variation in $\rho$ very soon ($t=10$) drops down to $10^{-5}$. In time, turbulence begins to take effect, and the variation is about $10^{-4}$ at about $t=300$ and $400$.

We now present the simulation with control. Recalling~\ref{subsec:control_strategy_2c}, the external control electric potential is
\begin{align*}
\hat{\Phi}(t,\veck) = - \hat{S} + c \delta(t) 
+ c \frac{\sin(B_0 t)}{B_0} \hat{\mu}(\tilde{\maxR}(t)\veck) \,.
\end{align*}
Setting $c=0$ in~\eqref{eq:control_lim_c} returns the Control Strategy 1. We will examine different effects when $c$ is set at different values.

\begin{figure}[htbp]
    \centering
    \begin{subfigure}[b]{0.32\textwidth}
        \centering
        \includegraphics[width=\textwidth]{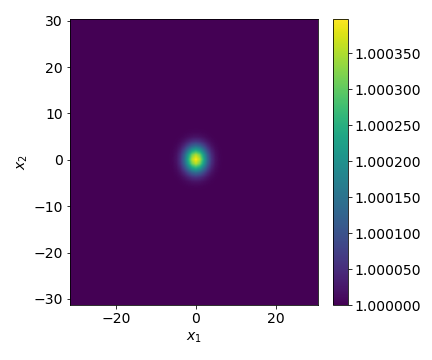}
        \caption{$t=0$}
    \end{subfigure}
    \hfill
    \begin{subfigure}[b]{0.32\textwidth}
        \centering
        \includegraphics[width=\textwidth]{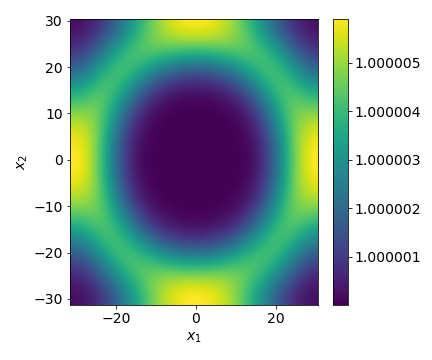}
        \caption{$t=20$}
    \end{subfigure}
    \hfill
    \begin{subfigure}[b]{0.32\textwidth}
        \centering
        \includegraphics[width=\textwidth]{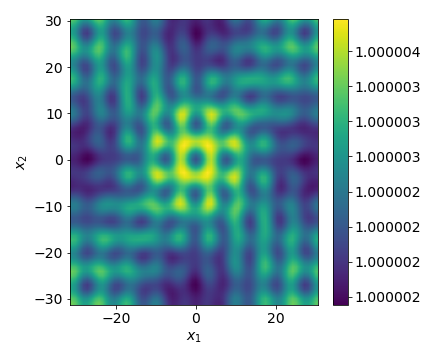}
        \caption{$t=50$}
    \end{subfigure}
    \\
    \begin{subfigure}[b]{0.32\textwidth}
        \centering
        \includegraphics[width=\textwidth]{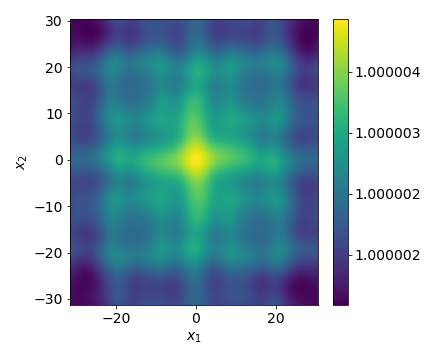}
        \caption{$t=200$}
    \end{subfigure}
    \hfill
    \begin{subfigure}[b]{0.32\textwidth}
        \centering
        \includegraphics[width=\textwidth]{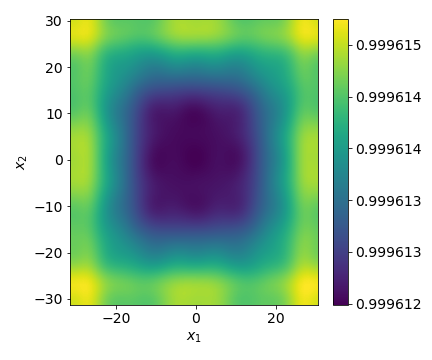}
        \caption{$t=300$}
    \end{subfigure}
    \hfill
    \begin{subfigure}[b]{0.32\textwidth}
        \centering
        \includegraphics[width=\textwidth]{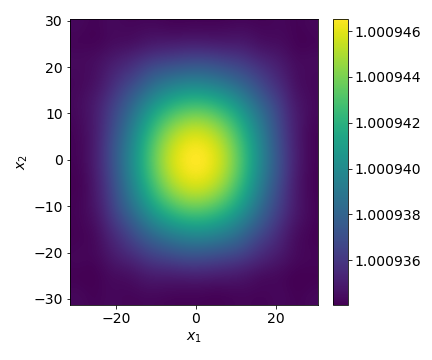}
        \caption{$t=400$}
    \end{subfigure}
    \caption{Density $\rho(x_1, x_2)$ snapshot at different time steps, with Control Strategy 1 that $c=0$.}
    \label{fig:DGH_gaussx_rhoxy_control_c0}
\end{figure}

\begin{figure}[htbp]
    \centering
    \begin{subfigure}[b]{0.24\textwidth}
        \centering
        \includegraphics[width=\textwidth]{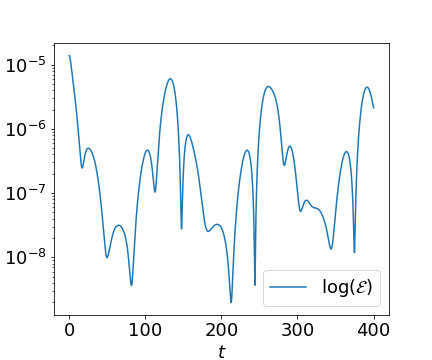}
        \caption{ }
    \end{subfigure}
    \hfill
    \begin{subfigure}[b]{0.24\textwidth}
        \centering
        \includegraphics[width=\textwidth]{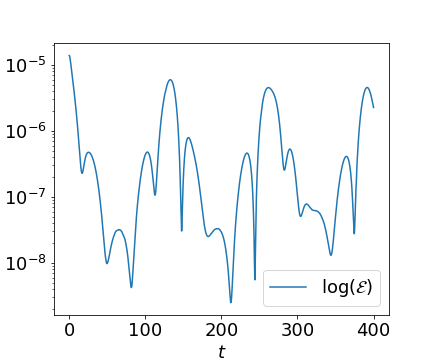}
        \caption{ }
    \end{subfigure}
    \hfill
    \begin{subfigure}[b]{0.24\textwidth}
        \centering
        \includegraphics[width=\textwidth]{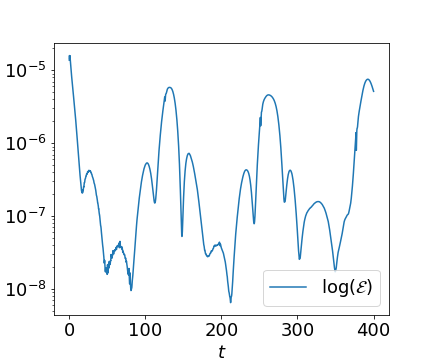}
        \caption{ }
    \end{subfigure}
    \hfill
    \begin{subfigure}[b]{0.24\textwidth}
        \centering
        \includegraphics[width=\textwidth]{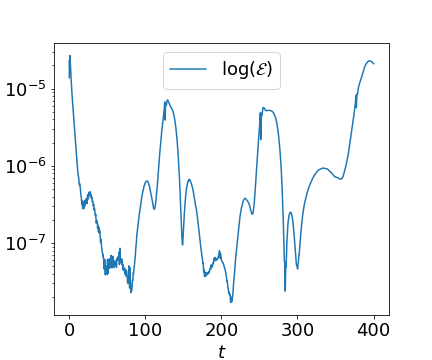}
        \caption{ }
    \end{subfigure}
    \\
    \begin{subfigure}[b]{0.24\textwidth}
        \centering
        \includegraphics[width=\textwidth]{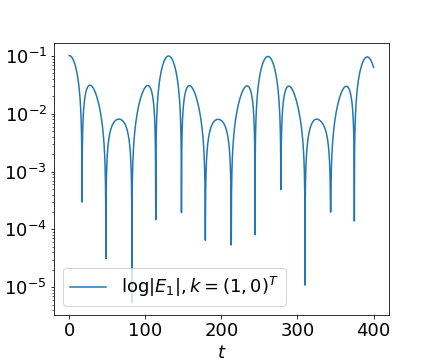}
        \caption{ }
    \end{subfigure}
    \hfill
    \begin{subfigure}[b]{0.24\textwidth}
        \centering
        \includegraphics[width=\textwidth]{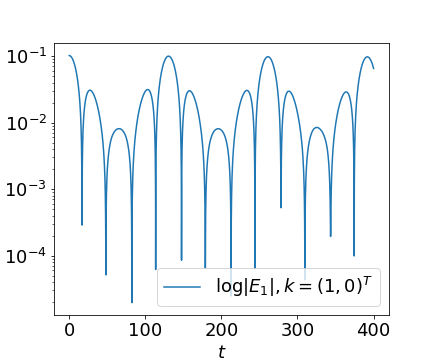}
        \caption{ }
    \end{subfigure}
    \hfill
    \begin{subfigure}[b]{0.24\textwidth}
        \centering
        \includegraphics[width=\textwidth]{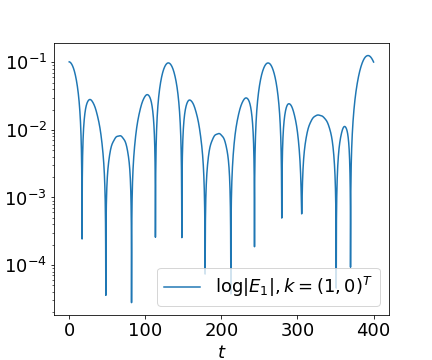}
        \caption{ }
    \end{subfigure}
    \hfill
    \begin{subfigure}[b]{0.24\textwidth}
        \centering
        \includegraphics[width=\textwidth]{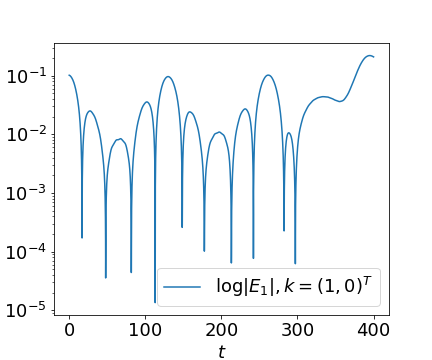}
        \caption{ }
    \end{subfigure}
    \caption{Evolution of the electric energy $\mathcal{E}(t)$, and the first coordinate of $|\vecE_{(1,0)^\top}(t)|$ in semi-log scale, control with different $c$ values. Panel (a)(e): $c=0$, Panel (b)(f): $c=0.001$, Panel (c)(g): $c=0.005$, Panel (d)(h): $c=0.01$.}
    \label{fig:DGH_gaussx_energy_control_various_c}
\end{figure}

When $c=0$, the corresponding electric energy and Fourier mode coefficient are presented in Figure~\ref{fig:DGH_gaussx_energy_control_various_c}(a)\revise{(e)}. The electric energy is suppressed under $10^{-5}$ for all time, and the density profile has much less variation comparing Figure~\ref{fig:DGH_gaussx_rhoxy_control_c0} to Figure~\ref{fig:DGH6_gaussx_nonlinear_rhoxy}.

We also tested setting $c$ to be at different values. The results are shown in Figure~\ref{fig:DGH_gaussx_energy_control_various_c}, with $c=0.001, c=0.005, c=0.01$. For all these tests, the electric energy can be controlled below the level of $10^{-5}$ for all time, thus suppressing the instability as in Figure~\ref{fig:DGH_gaussx_energy_no_control}. However, we also notice that in Figure~\ref{fig:DGH_gaussx_energy_control_various_c}\revise{(d)}(h), for $c$ being $0.01$, the electric energy gradually grows after $t=300$. This indicates that the perturbation induced by the term $c\frac{\sin(B_0 t)}{B_0}\hat{\mu}$ in~\eqref{eq:control2_c_formula} might lead to another instability in the system.

\subsection{Gaussian equilibrium}  
\begin{figure}[htbp]
    \centering
    \begin{subfigure}[b]{0.4\textwidth}
        \centering
        \includegraphics[width=\textwidth]{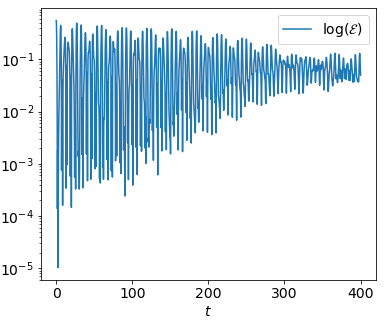}
        \caption{ }
    \end{subfigure}
    \hfill
    \begin{subfigure}[b]{0.4\textwidth}
        \centering
        \includegraphics[width=\textwidth]{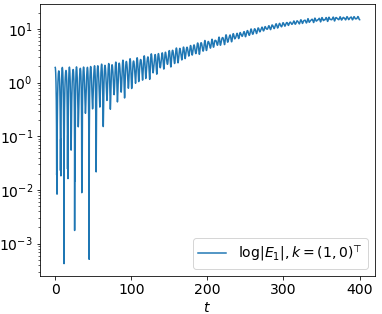}
        \caption{ }
    \end{subfigure}
    \caption{Evolution of the electric energy $\mathcal{E}(t)$, and the first coordinate of $|\vecE_{(1,0)^\top}(t)|$ in semi-log scale, without control.}
    \label{fig:nonlinear_Gaussian_energy_no_control}
\end{figure}

\begin{figure}[htbp]
    \centering
    \begin{subfigure}[b]{0.32\textwidth}
        \centering
        \includegraphics[width=\textwidth]{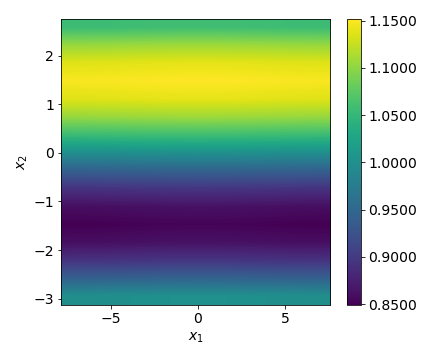}
        \caption{$t=0$}
    \end{subfigure}
    \hfill
    \begin{subfigure}[b]{0.32\textwidth}
        \centering
        \includegraphics[width=\textwidth]{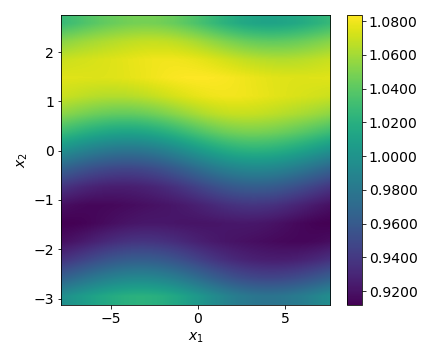}
        \caption{$t=200$}
    \end{subfigure}
    \hfill
    \begin{subfigure}[b]{0.32\textwidth}
        \centering
        \includegraphics[width=\textwidth]{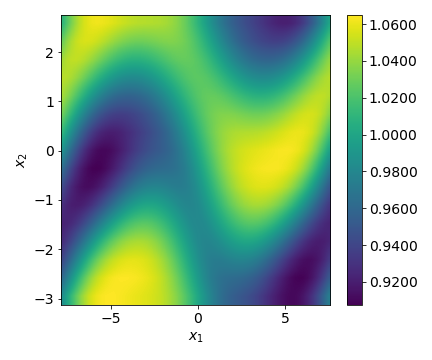}
        \caption{$t=300$}
    \end{subfigure}
    \\
    \begin{subfigure}[b]{0.32\textwidth}
        \centering
        \includegraphics[width=\textwidth]{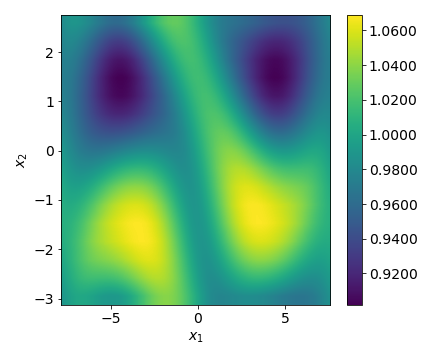}
        \caption{$t=350$}
    \end{subfigure}
    \hfill
    \begin{subfigure}[b]{0.32\textwidth}
        \centering
        \includegraphics[width=\textwidth]{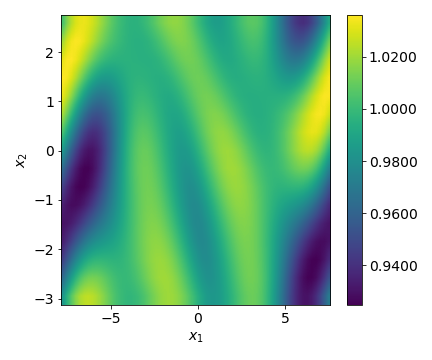}
        \caption{$t=375$}
    \end{subfigure}
    \hfill
    \begin{subfigure}[b]{0.32\textwidth}
        \centering
        \includegraphics[width=\textwidth]{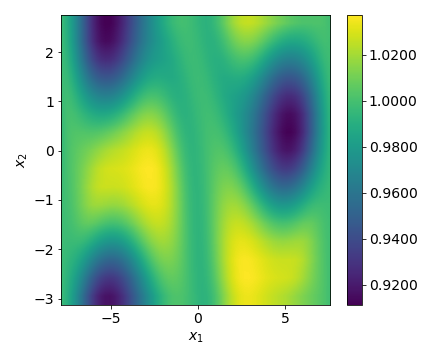}
        \caption{$t=400$}
    \end{subfigure}
    \caption{Density $\rho(x_1, x_2)$ snapshot at different time steps, without control.}
    \label{fig:Gauss_no_control_rhoxy}
\end{figure}

\begin{figure}[htbp]
    \centering
    \begin{subfigure}[b]{0.325\textwidth}
        \centering
        \includegraphics[width=\textwidth]{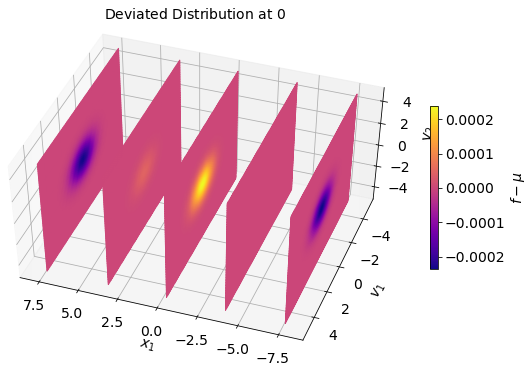}
        \caption{$t=0$}
    \end{subfigure}
    \hfill
    \begin{subfigure}[b]{0.325\textwidth}
        \centering
        \includegraphics[width=\textwidth]{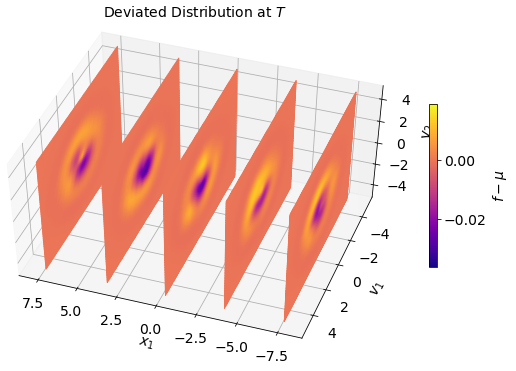}
        \caption{$t=400$}
    \end{subfigure}
    \hfill
    \begin{subfigure}[b]{0.325\textwidth}
        \centering
        \includegraphics[width=\textwidth]{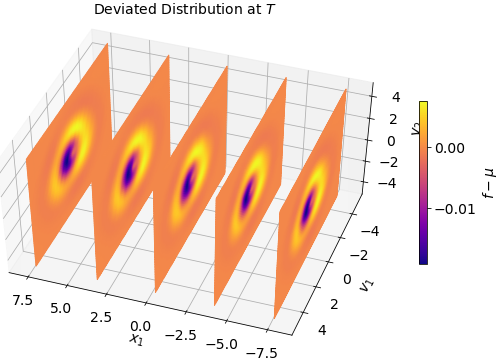}
        \caption{$t=400$}
    \end{subfigure}
    \caption{The deviated distribution function $f(x_1, x_2=0, v_1, v_2)-\mu$ at five $x_1$ locations. Panels (a) and (b) show the deviated distribution without control at time $t=0$ and $t=400$, respectively. Panels (a) and (c) show the deviated distribution with Control Strategy 1, that $c=0$.}
    \label{fig:Gaussian_f_dist_no_control}
\end{figure}

Gaussian equilibrium is expected to be a linearly stable equilibrium state according to the linear analysis when the initial perturbation is small. We now study the situation when a large perturbation is posed on the Gaussian equilibrium. This is beyond the linear stable regime, and potential instability may be observed. This is indeed the case. We set $B_0=1$, adopt the Kelvin-Helmholtz type instability example in~\cite{vu:hal-04016348} and use the following initial condition:
\begin{align*}
F(t=0, \vecx, \vecv) = (1+ 0.0015 \cos(0.4x_1) + 0.15 \sin(x_2))\frac{1}{2\pi} e^{-\frac{v_1^2 + v_2^2}{2}}\,,
\end{align*}
where $x_1 \in [-2.5\pi, 2.5\pi), x_2 \in [-\pi, \pi), v_1, v_2\in[-5,5].$ For numerical discretization, we use $N_{x_1}=64, N_{x_2}=16, N_{v_1}=N_{v_2}=64$ and run the simulation till terminal time $T=400$. Without any control, we present the electric energy, together with the Fourier coefficient $|\vecE_{(1,0)^\top}|$ in Figure~\ref{fig:nonlinear_Gaussian_energy_no_control}. The specific feature of Bernstein modes is still evident, in the sense that the electric energy is bounded and oscillatory, i.e., no damping. However, exponential growth is observed for certain Fourier modes, such as $\veck=(1,0)^\top$ in Panel (b). We also plot six time slots of the density $\rho$ in Figure~\ref{fig:Gauss_no_control_rhoxy}. According to the initial distribution formula, the initial density is in the shape of a sine wave in the $x_2$ direction, but the perturbation in the $x_1$ direction is negligible. However, as the dynamics continues, perturbations in the $x_1$ direction become more and more noticeable. A full 3D distribution (set $x_2=0$) at initial time $t=0$ and final time $t=400$ is shown in Figure~\ref{fig:Gaussian_f_dist_no_control} (a)(b).

\begin{figure}[htbp]
    \centering
    \begin{subfigure}[b]{0.4\textwidth}
        \centering
        \includegraphics[width=\textwidth]{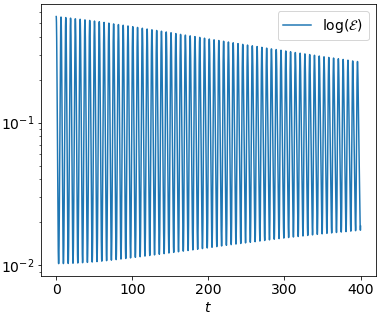}
        \caption{ }
    \end{subfigure}
    \hfill
    \begin{subfigure}[b]{0.44\textwidth}
        \centering
        \includegraphics[width=\textwidth]{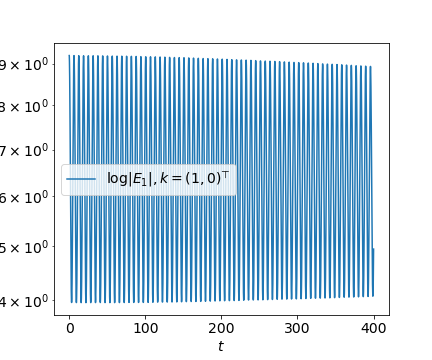}
        \caption{ }
    \end{subfigure}
    \caption{Evolution of the electric energy $\mathcal{E}(t)$, and the first coordinate of $|\vecE_{(1,0)^\top}(t)|$ in semi-log scale, with Control Strategy 1 that $c=0$.}
    \label{fig:nonlinear_Gaussian_energy_pole_control}
\end{figure}

\begin{figure}[htbp]
    \centering
    \begin{subfigure}[b]{0.32\textwidth}
        \centering
        \includegraphics[width=\textwidth]{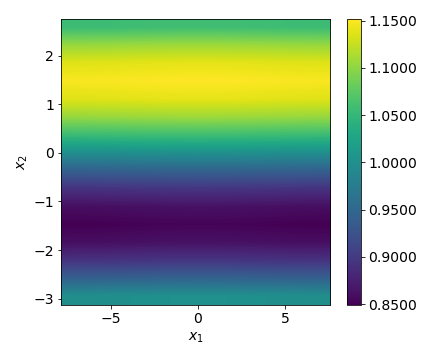}
        \caption{$t=0$}
    \end{subfigure}
    \hfill
    \begin{subfigure}[b]{0.32\textwidth}
        \centering
        \includegraphics[width=\textwidth]{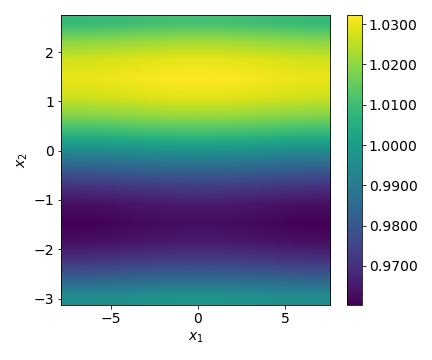}
        \caption{$t=300$}
    \end{subfigure}
    \hfill
    \begin{subfigure}[b]{0.32\textwidth}
        \centering
        \includegraphics[width=\textwidth]{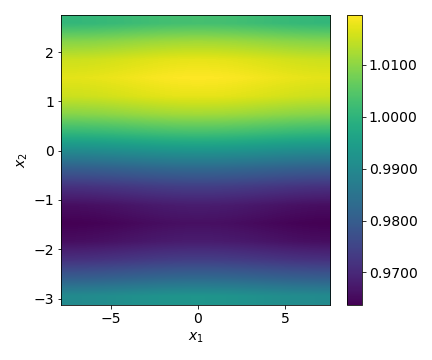}
        \caption{$t=400$}
    \end{subfigure}
    \caption{Density $\rho(x_1, x_2)$ snapshot at different time steps, with Control Strategy 1 that $c=0$.}
    \label{fig:Gauss_control_rhoxy}
\end{figure}

We then apply Control Strategy 1 to the above system. According to Remark~\ref{rmk:nonlinear_stream}, the control is expected to be effective, as will be shown below in numerical results. In Figure~\ref{fig:nonlinear_Gaussian_energy_pole_control}, it is clear that the electric energy and $\veck=(1,0)^\top$ continue being oscillatory, but the magnitude decays in time, and thus is suppressed. The density profile is also much smoother, as seen in Figure~\ref{fig:Gauss_control_rhoxy}. It is kept close to the initial perturbation for all time, and there is no mixing compared to Figure~\ref{fig:Gauss_no_control_rhoxy}. According to Figure~\ref{fig:Gaussian_f_dist_no_control} (a)(c), where we present the initial and final 3D distribution (with fixed $x_2=0$), it is also clear that turbulence no longer exists.

\section{Conclusion \label{sec:conclusion}}
This paper studies the stabilization and control of plasma dynamics in the presence of an external magnetic field, modeled by the magnetized Vlasov–Poisson system. We derived the corresponding dispersion relation and formulated the associated Penrose-type stability condition (Proposition~\ref{prop:penrose_cond}). As representative examples, we considered two equilibrium distributions: a Gaussian equilibrium and a ring-shaped equilibrium. The latter admits an unstable root in the dispersion relation, leading to linear instability (the Dory–Guest–Harris instability) and exponential growth of the electric energy in the linear regime.

Building on this analysis, we investigated the control problem of suppressing instability through an externally imposed electric field. The dispersion relation provides a central tool for the design of such controls. We proposed two control strategies, both based on eliminating unstable roots in the dispersion relation. One of these strategies recovers the free-streaming solution as a special case, which further enables an explicit characterization of the electric energy and a proof that it remains bounded. Numerical experiments demonstrate the effectiveness of the proposed strategies and indicate that stabilization persists beyond the strictly linear regime, including in certain nonlinear settings.

\revise{Stabilizing plasma dynamics via external fields is a key component in the broader objective of controlled fusion. While the present work is carried out in an idealized setting, it provides a step toward systematic control design in higher-dimensional kinetic models relevant to fusion applications. An important future direction is to extend the pole-elimination methodology to reduced descriptions such as drift-kinetic and gyrokinetic regimes, where the presence of multiple time and length scales introduces additional analytical and computational challenges. On the computational side, advancing to gyrokinetic scaling will likely require efficient large-scale solvers, potentially accelerated by GPU-based implementations.}

\backmatter

\subsection*{Acknowledgements}
P.C. thanks Genia Vogman, Xiaochuan Tian, and Qihao Ye for useful discussions on the Dory-Guest-Harris instability and related details.
R.J. is supported by the National Science Foundation under Grant No. 2409066.
P.C., Q.L., and Y.Y. are supported by the National Science Foundation under Grant No.2308440.

\section*{Declarations}

The authors have no competing interests to declare that are relevant to the content of this article.

\begin{appendices}

\section{Computation of 
\texorpdfstring{$P(\veck, \lambda)$}{Pklambda}}
\subsection{Fourier transform of ring-shaped \texorpdfstring{$\mu$}{mu} \texorpdfstring{\eqref{eqn: DGH-fv}}{mueqn} }\label{appx:DGH_Fourier_compute}
In this subsection, we present the details to compute the Fourier transform of the ring-shaped distribution~\eqref{eqn: DGH-fv}.

Note that $\mu$ is radial symmetric in $\vecv$, we change to polar coordinates, then
\begin{align*}
\hat{\mu}(\veck) &= \frac{1}{\pi \alpha_{\perp}^2 j!} \int_{\mathbb{R}^2} e^{-i (k_1 v_1 + k_2 v_2)}  \left(\frac{v_1^2 + v_2^2}{\alpha_{\perp}^2}\right)^j \exp{\left(-\frac{v_1^2 + v_2^2}{\alpha_{\perp}^2}\right)}\,d\vecv\,,\\
&= \frac{1}{\pi \alpha_{\perp}^2 j!} \frac{1}{\alpha_{\perp}^{2j}} \int_{0}^{\infty} \int_0^{2\pi} e^{-i (k_1 v_r \cos\theta + k_2 v_r \sin\theta)}  v_r^{2j} \exp{\left(-\frac{v_r^2}{\alpha_{\perp}^2}\right)}\,v_r dv_r d\theta\,,\\
& = \frac{1}{\pi j!} \int_{0}^{\infty} \int_0^{2\pi} e^{-i \alpha_{\perp} w_r |\veck|\sin(\theta)} d\theta\, w_r^{2j+1} \exp{\left(-w_r^2\right)}\, dw_r \,, & (w_r := \frac{v_r}{\alpha_{\perp}})\\
& = \frac{2}{j!} \int_{0}^{\infty} J_0(\alpha_{\perp} |\veck| w_r)  w_r^{2j+1} \exp{\left(-w_r^2\right)}\, dw_r \,,
\end{align*}
where $J_0(\cdot)$ is the zeroth-order Bessel function of the first kind, and we use the integral representation of Bessel functions. Thus in~\eqref{eq:penrose_Pk_lam},
\begin{align*}
\hat{\mu}(\tilde{\maxR}\veck) & = \frac{2}{j!} \int_{0}^{\infty} J_0(\alpha_{\perp} |\tilde{\maxR} \veck| w_r)  w_r^{2j+1} \exp{\left(-w_r^2\right)}\, dw_r \,,\\
& = M(j+1,1,-\frac{\alpha_{\perp}^2 |\tilde{\maxR} \veck|^2}{4}) = \exp{\left(-\frac{\alpha_{\perp}^2 |\tilde{\maxR} \veck|^2}{4}\right)} M(-j,1,\frac{\alpha_{\perp}^2 |\tilde{\maxR} \veck|^2}{4})\,,
\end{align*}
here $M(\cdot,\cdot,\cdot)$ is the Kummer's function (or the hypergeometric function), the last line is obtained in~\cite{integral-Gradshteyn-Ryzhik-2014} together with Kummer's transform. 
One specific property of Kummer's function is that for $j$ being nonnegative integers, $M(-j,\cdot,\cdot)$ can be expressed as a polynomial of degree no more than $j$.

In the case of $j=6$~\eqref{eq:DGH6_equilibrium}, the analytical expression of $M$ is a polynomial of power $6$:
\begin{align*}
M(-6,1,z) =& \sum_{n=0}^6 \frac{(-1)^n 6!/n!}{(n!)^2} z^n
= 1 -6z + \frac{15}{2}z^2 - \frac{10}{3} z^3 + \frac{5}{8} z^4 - \frac{1}{20}z^5 + \frac{1}{6!} z^6\,,
\end{align*}
where $z=\frac{\alpha_{\perp}^2 |\tilde{\maxR} \veck|^2}{4}=\frac{1}{6 B_0^2}|\veck|^2(1-\cos(B_0 t))$.

\end{appendices}

\bibliography{000sn-bibliography_new} 

\end{document}